\documentclass[12pt,subeqn,a4paper]{article}

\usepackage{amssymb,amsmath,amsfonts,amsthm, amscd, mathrsfs,helvet}
\usepackage{bm}
\usepackage{graphicx,verbatim}
\usepackage{psfrag}
\usepackage[all]{xy}

\theoremstyle{plain}

\newcommand{\remark}[1]{\fontsize{8pt}{5pt}\selectfont #1}

\numberwithin{equation}{section}

\begin{document}

\begin{titlepage}
\begin{flushright}
{\bf June 2006} \\
DAMTP-\\

hep-th/0606287 \\
\end{flushright}
\begin{centering}
\vspace{.2in}
{\large {\bf A Symplectic Structure for String Theory \\
on Integrable Backgrounds}}\\

\vspace{.3in}

Nick Dorey and Beno\^{\i}t Vicedo\\
\vspace{.1 in}
DAMTP, Centre for Mathematical Sciences \\
University of Cambridge, Wilberforce Road \\
Cambridge CB3 0WA, UK \\
\vspace{.2in}
%
%
\vspace{.4in}

{\bf Abstract} \\
\end{centering}

We define regularised Poisson brackets for the monodromy matrix of
classical string theory on $\mathbb{R} \times S^3$. The
ambiguities associated with Non-Ultra Locality are resolved using
the symmetrisation prescription of Maillet. The resulting brackets
lead to an infinite tower of Poisson-commuting conserved charges
as expected in an integrable system. The brackets are also used to
obtain the correct symplectic structure on the moduli space of
finite-gap solutions and to define the corresponding action-angle
variables. The canonically-normalised action variables are the
filling fractions associated with each cut in the finite-gap
construction. Our results are relevant for the leading-order
semiclassical quantisation of string theory on $AdS_{5}\times
S^{5}$ and lead to integer-valued filling fractions in this
context.

\end{titlepage}

\setcounter{page}{0}

\tableofcontents

\input{epsf}

\setcounter{section}{-1}

\section{Introduction}
\begin{figure}
\centering \psfrag{a}{$\mathcal{A}_I$} \psfrag{b}{$\mathcal{B}_I$}
\psfrag{c}{$\mathcal{C}_I$} \psfrag{pinf}{$\infty^+$}
\psfrag{minf}{$\infty^-$} \psfrag{pp}{$p(x)$} \psfrag{mp}{$-p(x)$}
\includegraphics{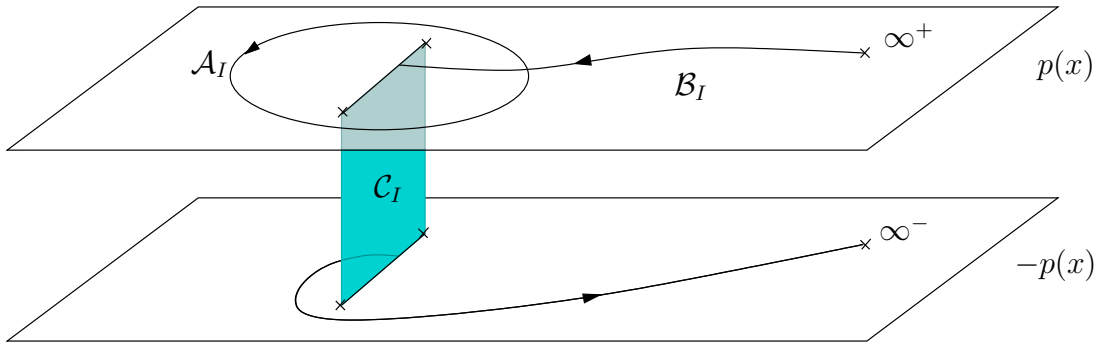}
\caption{The cycle $\mathcal{A}_I$ and path $\mathcal{B}_I$ for
the cut $\mathcal{C}_I$.} \label{AnBcycles}
\end{figure}

\paragraph{}
Determining the exact spectrum of free string theory on
$AdS_{5}\times S^{5}$ is an important problem whose solution would
surely lead to a better understanding of the AdS/CFT
correspondence. The discovery of integrability in the classical
theory \cite{Bena:2003wd} is a good indication that the problem
may be tractable. More precisely, the authors of
\cite{Bena:2003wd} found a Lax formulation of the equations of
motion which leads the existence of an infinite tower of conserved
charges in the classical worldsheet theory. These charges have
subsequently been exploited to construct and classify large
families of exact solutions of the classical equations of motion
\cite{Paper1, Kazakov:2004qf, Beisert:2004ag,
Beisert:2005bm, Schafer-Nameki:2004ik,Alday:2005gi}.
However, this does not quite coincide with the standard definition
of integrability. Integrability in the standard sense requires not
only the existence of a tower of conserved charges but also
requires that these charges be ``in involution''. In other words
the conserved charges should Poisson commute with each other. In
finite-dimensional systems, this is a necessary condition for
Liouville's theorem\footnote{Liouville's theorem applies to
dynamical systems with $N$ degrees of freedom which also have $N$,
globally defined, conserved charges in involution. The theorem
guarantees that the equations of motion can be solved by
quadratures for arbitrary initial data (see e.g. \cite{Babelon}).}
to hold. More generally, knowledge of the Poisson brackets is
necessary for constructing the action-angle variables for the
system which play a key role in semiclassical quantisation. In
this paper, which builds on our earlier work \cite{Paper1}, we
will derive the involution condition for classical strings moving
on an $\mathbb{R}\times S^{3}$ submanifold of $AdS_{5}\times
S^{5}$ and construct the corresponding action-angle variables.
\paragraph{}
In classical string theory on $AdS_{5}\times S^{5}$, as well as
many other backgrounds which admit a Lax formulation, there is a
long-standing problem in determining the Poisson brackets of the
conserved charges. As we review below, the problem is due to the
presence of {\em Non-Ultra Local} (NUL) terms in the Poisson
brackets of the worldsheet fields which lead to ambiguities in
brackets for the charges. In this paper we will present a
resolution of this problem based on earlier work by Maillet
\cite{Maillet, Maillet2, Maillet3} in the context of two
dimensional field theory. In particular, Maillet proposed a
prescription for regularising the problematic brackets. In the
following we will apply his procedure to the simplest classical
subsector of the $AdS_5 \times S^5$ theory which corresponds to
bosonic strings moving on an $\mathbb{R} \times S^3$ submanifold
of the full geometry. We will show that this prescription leads to
a very natural symplectic structure on the space of finite-gap
solutions of the string equations of motion constructed in
\cite{Paper1}. In particular, we find that this symplectic
structure leads to canonically normalised action variables which
are exactly equal to the filling fractions discussed in
\cite{Paper1}. Our results are relevant for the leading-order
semiclassical quantisation of strings moving on an
$\mathbb{R}\times S^{3}$ submanifold of $AdS_{5}\times S^{5}$. In
this context, they confirm the expected integer quantisation of
the filling fractions discussed in \cite{Paper1}. Our methods
should generalise to other sectors of classical strings on
$AdS_{5}\times S^{5}$ and also to other integrable backgrounds. In
the rest of this introductory section we will outline the main
ideas in the paper.
\paragraph{}
Bosonic strings moving on $\mathbb{R} \times S^3$ are described in
static gauge by an $SU(2)$-valued world-sheet field
$g(\sigma,\tau)$ which gives rise to a conserved current
$j_{\mu}(\sigma,\tau)=-g^{-1}\partial_{\mu}g$. The corresponding
action for $g(\sigma,\tau)$ is essentially that of the $SU(2)$ Principal
Chiral model,
\begin{equation}
S=\frac{\sqrt{\lambda}}{4\pi}\int \,d\sigma d\tau\,\,\frac{1}{2}{\rm
  tr}(j_{\mu}j^{\mu})
\label{principal}
\end{equation}
where $\lambda$ is a dimensionless coupling constant. Physical motions
of the string also obey the Virasoro constraint,
\begin{equation} \label{V}
\frac{1}{2} \text{tr} j_{\pm}^2 = -\kappa^2
\end{equation}
where $j_{\pm}=j_{0}\pm j_{1}$ are the lightcone components of the
current and $\kappa$ is a constant related to the spacetime energy
of the string. For many purposes it is convenient to complexify
the model and work with a current $j_{\mu}$ taking values in the
Lie algebra $\mathfrak{sl}(2,{\mathbb{C}})$. A solution of the
original problem where $j_{\mu}$ is restricted to lie in
$\mathfrak{su}(2)$ is then obtained by imposing appropriate
reality conditions.
\paragraph{}
Starting from the action (\ref{principal}) it is straightforward
to obtain the (equal-$\tau$) Poisson brackets for the components
of the current $j_{\mu}(\sigma)$. Writing the current as $j_0 =
j_0^a t^a$, $j_1 = j_1^a t^a$, in terms of $SU(2)$ generators $t^a$
satisfying,
\begin{equation*}
[t^a,t^b] = f^{abc} t^c, \quad \text{tr}(t^a t^b) = - \delta^{ab}.
\end{equation*}
the resulting brackets are,
\begin{equation} \label{PB0}
\begin{split}
\left\{ j_1^a(\sigma), j_1^b(\sigma')\right\} &=  0, \\
\frac{\sqrt{\lambda}}{4 \pi} \left\{ j_0^a(\sigma),
j_1^b(\sigma')\right\} &=  - f^{abc} j_1^c(\sigma) \delta(\sigma -
\sigma') - \delta^{ab}
\delta'(\sigma - \sigma'), \\
\frac{\sqrt{\lambda}}{4 \pi} \left\{ j_0^a(\sigma),
j_0^b(\sigma')\right\} &= - f^{abc} j_0^c(\sigma) \delta(\sigma -
\sigma').
\end{split}
\end{equation}
\paragraph{}
These brackets are usually described as Non-Ultra Local (NUL)
reflecting the presence the the derivative $\delta'(\sigma-\sigma')$
in the second bracket. As we now review, the problems related
to the NUL nature of these brackets emerge when we consider the corresponding
Poisson brackets of the infinite tower of conserved charges of the
model. The starting point for constructing these charges is
the existence of a one-parameter family of flat currents,
\begin{equation} \label{LC}
J(x) = \frac{1}{1 - x^2} (j - x \ast j),
\end{equation}
labelled by the complex spectral parameter $x\in \mathbb{C}$. The
flatness of $J(x)$, for all values of $x$, is equivalent to the
equations of motion which follow from the action (\ref{principal}).
\paragraph{}
Using the current $J(x)$, we can construct a monodromy matrix,
\begin{equation} \label{MD}
\Omega(x,\sigma,\tau) = P \overleftarrow{\exp}
\int_{[\gamma(\sigma,\tau)]} J(x)\,\,\in\,\, SL(2,\mathbb{C})
\end{equation}
where $\gamma(\sigma,\tau)$ is a non-contractible loop on the
string worldsheet based at the point $(\sigma,\tau)$. The flatness
of $J(x)$ implies that $\Omega(x)$ undergoes isospectral evolution
in the world-sheet coordinates. In other words the eigenvalues of
the monodromy matrix are independent of $\sigma$ and $\tau$. As
$\Omega(x)$ takes values in $SU(2)$ when $x \in \mathbb{R}$, it is
convenient to parametrise the eigenvalues as,
\begin{equation}
\lambda_{\pm}=\exp\left(\pm ip(x)\right).
\label{quasi1}
\end{equation}
Here $p(x)$ is a (multi-valued) function of the spectral parameter
which is known as the {\em quasi-momentum}. The Taylor coefficients in
the expansion of $p(x)$ then generate an infinite tower of
conserved quantities on the worldsheet.
\paragraph{}
The Poisson bracket for the conserved charges can be deduced from
the Poisson bracket $B(x,x')=\{\Omega(x)\, \overset{\otimes}, \,
\Omega(x')\}$ for the monodromy matrix.
To calculate $B(x,x')$, we begin by defining a
transition matrix between distinct points $\sigma_{1}$ and
$\sigma_{2}$ on the string,
\begin{equation*}
T(\sigma_1,\sigma_2,x) = P \overleftarrow{\exp}
\int_{\sigma_2}^{\sigma_1} d\sigma J_1(\sigma,x).
\end{equation*}
Using the Poisson brackets (\ref{PB0}) of the current we can calculate
the bracket,
\begin{equation*}
\Delta^{(1)}(\sigma_1,\sigma_2,\sigma'_1,\sigma'_2; x,x') = \{
T(\sigma_1,\sigma_2,x) \, \overset{\otimes}, \,
T(\sigma'_1,\sigma'_2,x') \}
\end{equation*}
This is well defined when the points $\sigma_1, \sigma_2, \sigma'_1,
\sigma'_2$ are all distinct. However the presence of the distribution
$\delta'(\sigma-\sigma')$ on the RHS of the second bracket in
(\ref{PB0}) leads to a finite discontinuity on surfaces where two of
the points coincide. To obtain the desired bracket $B(x,x')$
we must take the limit $\sigma_1\rightarrow  \sigma'_1$,
$\sigma_2\rightarrow \sigma'_2$ and the discontinuity of
$\Delta^{(1)}$ on this surface leads to an ambiguous result.
\paragraph{}
The ambiguity described above is quite mild for the bracket
$B(x,x')$ itself, but becomes more serious when one tries to
define nested Poisson brackets for a product of monodromy
matrices. The ambiguities then result from multiple coincident
endpoints of the corresponding transition matrices. To resolve the
ambiguity one can introduce an infinitesimal splitting between
these coincident endpoints. Fortunately there is a straightforward
prescription due to Maillet which seems to provide the unique
consistent resolution of the problem. As we review in Section
\ref{section: Maillet}, Maillet's prescription involves a total
symmetrisation over all possible point-splittings. The
prescription preserves the defining properties of the Poisson
bracket such as its anti-symmetry, the Leibniz rule and the Jacobi
identity. The resulting bracket of two monodromy matrices can be
then be written as,
\begin{align} \label{fundamental Poisson bracket0}
\left\{ \Omega(x) \mathop{,}^{\otimes} \Omega(x') \right\} =
&[r(x,x'), \Omega(x) \otimes \Omega(x')] \notag \\ +
&\left(\Omega(x) \otimes {\bf 1}\right) s(x,x') \left( {\bf 1}
\otimes \Omega(x') \right) \notag \\ - &\left( {\bf 1} \otimes
\Omega(x') \right) s(x,x') \left( \Omega(x) \otimes {\bf 1} \right),
\end{align}
where,
\begin{equation} \label{rs-matrix0}
r(x,x') = - \frac{2 \pi}{\sqrt{\lambda}} \frac{x^2 + {x'}^2 - 2 x^2
{x'}^2}{(x-x')(1 - x^2)(1 - {x'}^2)}, \qquad s(x,x') = - \frac{2
\pi}{\sqrt{\lambda}} \frac{x+x'}{(1 - x^2)(1 - {x'}^2)}.
\end{equation}
Finally, using this relation one may compute the bracket,
\begin{align} \label{involution of conserved quantities0}
\left\{ \text{tr }\Omega(x)^n, \text{tr }\Omega(x')^m \right\} = 0.
\end{align}
As above the eigenvalues of $\Omega(x)$ yield a one-parameter
family of conserved charges. The bracket \eqref{involution of
conserved quantities0} therefore implies that the charges
corresponding to different values of the spectral parameter $x$
Poisson commute. This is the natural generalisation of the
involution condition discussed above for an infinite dimensional
system.
\paragraph{}
The main goal of this paper is to explore the consequence of
Maillet's prescription for the finite-gap solutions of the string
equations of motion discussed in \cite{Paper1}. Solutions carry
the conserved charges $Q_{L}$ and $Q_{R}$ associated with the
$SU(2)_{L}\times SU(2)_{R}$ isometry group of the target $S^{3}$.
As in \cite{Paper1} we will focus on solutions of highest weight
with respect to both $SU(2)$ factors which have,
\begin{equation*}
Q_R = \frac{1}{2i}R \sigma_3, \; Q_L = \frac{1}{2i}L \sigma_3
\end{equation*}
where $\sigma_3 = \text{diag}(1,-1)$ is the third Pauli matrix.
The required solutions are characterised by the analytic behaviour
of the corresponding quasi-momentum $p(x)$ in the spectral plane.
The definition (\ref{quasi1}) implies that $p(x)$ need not be
single-valued, but can have discontinuities of the form
\begin{equation}\label{mode}
p(x+\epsilon) + p(x-\epsilon)=2\pi n_I, \quad x\in \mathcal{C}_I, \quad
n_I \in \mathbb{Z}, \quad I = 1,\ldots,K.
\end{equation}
across square-root branch cuts $\mathcal{C}_{I}$ in the $x$-plane. Finite-gap
solutions correspond  the case where $K$, the number of such cuts, is
finite. In this case, the
resulting double-cover of the $x$ plane defines a hyperelliptic
Riemann surface $\Sigma$ of finite genus $g=K-1$ known as the spectral
curve. It is convenient to define a basis of one-cycles on $\Sigma$ as
follows. For $I=1,\ldots,K$, the contour $\mathcal{A}_I$ surrounds the cut
$\mathcal{C}_I$ on the upper sheet while $\mathcal{B}_I$ runs from
the point at infinity on the upper sheet to the same point on the
lower sheet via the cut $\mathcal{C}_I$ (see Figure \ref{AnBcycles}).
The quasi-momentum $p(x)$ gives rise to a meromorphic differential $dp$ on
$\Sigma$ with periods,
\begin{equation} \label{AnB periods of dp}
\int_{\mathcal{A}_I} dp = 0, \quad \int_{\mathcal{B}_I} dp = 2 \pi
n_I, \; n_I \in \mathbb{Z}.
\end{equation}
\paragraph{}
The explicit reconstruction of solutions from the holomorphic data
$\{ \Sigma, dp \}$ was described in detail in \cite{Paper1} and is
also reviewed below in Section \ref{section: finite-gap}. Here we
will only summarise the main features. After taking into account
the various constraints on the data, Riemann surfaces $\Sigma$ and
differentials $dp$ corresponding to physical solutions are
parametrised by $K$ moduli $\mathcal{S}_{I}$ defined as,
\begin{equation}\label{FF}
\mathcal{S}_I = \frac{1}{2\pi i}\, \frac{\sqrt{\lambda}}{4\pi}\,
\int_{\mathcal{A}_I}\, \left(x+\frac{1}{x}\right)\, dp
\end{equation}
for $I=1,\ldots,K$. For real solutions, the moduli are real numbers
corresponding to the independent conserved charges of the model
carried by the configuration. They are further constrained by the
relations,
\begin{equation}\label{sumrule}
\sum_{I=1}^K \mathcal{S}_I = \frac{1}{2}(L - R), \qquad
\sum_{I=1}^K n_I \mathcal{S}_I = 0.
\end{equation}
The first equality suggests that we should identify $\mathcal{S}_{I}$
as the amount of angular momentum $J_{2}=(L-R)/2$ associated with each cut
$\mathcal{C}_{I}$. In the context of the AdS/CFT correspondence these
variables correspond to the filling fractions which count the total
number of Bethe roots associated with each cut. The second equation in
(\ref{sumrule}) corresponds to the constraint that the total
worldsheet momentum should vanish.
\paragraph{}
The moduli $\mathcal{S}_I$ correspond to the conserved quantities
of the corresponding string motion or the `action' variables. On
general grounds, we expect that each conserved quantity has a
corresponding conjugate variable which is periodic and evolves
linearly in time. The extra information required to uniquely
specify a solution is just the initial values of these `angle'
variables. In \cite{Paper1}, we identified this data with a
divisor $\gamma$ of degree $g$ on $\Sigma$ and an additional
angular variable $\bar{\theta}$ describing the global orientation
of the string. Here we will use an equivalent description in terms
of a divisor $\hat{\gamma}$ of degree $K=g+1$ on $\Sigma$. This in
turn uniquely specifies a point $\vec{\mathcal{A}}(\hat{\gamma})$ in the
generalised Jacobian $J(\Sigma,\infty^{\pm})$ (topologically
equivalent to $J(\Sigma)\times \mathbb{C}^{\ast}$) via the extended Abel map.
The $\sigma$ and $\tau$-evolution of the solution correspond to
the linear motion of this point. Finally to obtain a real
solution, the point $\vec{\mathcal{A}}(\hat{\gamma})$ is constrained to
lie on the real slice\footnote{See Section \ref{section:
finite-gap integration} and \ref{section: reality} for a more
precise discussion of the reality conditions.},
\begin{equation*}
T^K \simeq \text{Re} \left[ J(\Sigma)\times \mathbb{C}^{\ast} \right].
\end{equation*}
We define a set of coordinates
$\vec{\varphi}=(\varphi_1,\ldots,\varphi_K)$ on the real torus
$T^K$ normalised so that $\varphi_I \in [0,2\pi]$ for
$I=1,\ldots,K$. As these variables evolve linearly in the worldsheet
time they correspond to the normalised angle variables of the solution.
\paragraph{}
The space of finite-gap solutions is a real manifold of dimension
$2K$ parametrised by the coordinates $\{ S_I, \varphi_I \}_{I=1}^K$,
introduced above. The symplectic structure on the infinite
dimensional field space of the string defined by the regularised
Poisson brackets (\ref{fundamental Poisson bracket0}) induces a
symplectic structure on this manifold.
Our main result is an explicit formula for the corresponding
symplectic form $\hat{\omega}_{2K}$;
\begin{equation}
\hat{\omega}_{2K}= \sum_{I=1}^K \delta\mathcal{S}_I \wedge \delta
\varphi_I.
\label{symp0}
\end{equation}
As the angular variables $\varphi_{I}$ each have period $2\pi$, the
canonically conjugate variables $\mathcal{S}_{I}$ are the correctly
normalised action variables for the problem. The Bohr-Sommerfeld
condition for leading-order semiclassical quantisation of the
finite-gap solutions therefore simply imposes the integrality of the
filling fractions.
\paragraph{}
The rest of the paper is organised as follows. In Section
\ref{section: classical strings}, we describe the Hamiltonian
formulation of classical string theory on $\mathbb{R} \times S^3$.
In particular we derive the Poisson brackets \eqref{PB0} and
discuss the Maillet regularisation prescription leading to the
involution condition \eqref{involution of conserved quantities0}.
In Section \ref{section: data}, we use the regularised brackets to
obtain the symplectic form \eqref{symp0} on the space of
finite-gap solutions. Along the way, in subsection \ref{section:
finite-gap} we provide a review of the construction of finite-gap
solutions given in \cite{Paper1}. This subsection also contains a
new explicit formula for the original $\sigma$-model fields
corresponding to a genus $g$ finite-gap solution. Throughout this
section we emphasise the parallels between finite-gap solutions
and the conventional mode expansion for strings in flat space. The
remainder of Section \ref{section: data} describes the pullback of
the symplectic form to the moduli space of finite-gap solutions
and the corresponding action-angle variables. Some of the more
lengthy calculations are relegated to four Appendices.

\section{Classical integrability of strings on $\mathbb{R} \times
S^3$} \label{section: classical strings}

\subsection{Strings on $\mathbb{R} \times S^3$} \label{section: strings on S^3}

The embedding of the string in $\mathbb{R} \times S^3$ is
described by the time coordinate $X_0(\sigma,\tau) \in \mathbb{R}$
along with a matrix
\begin{equation} \label{g components}
g(\sigma,\tau) = \left( \begin{array}{cc} X_1 + i X_2 & X_3 + i X_4 \\
- X_3 + i X_4 & X_1 - i X_2 \end{array} \right) \equiv \left( \begin{array}{cc} Z_1 & Z_2\\
- \bar{Z}_2 & \bar{Z}_1 \end{array} \right) \in SU(2)
\end{equation}
describing the embedding in $S^3$. In conformal gauge the action
can be written in terms of a current $j = - g^{-1} dg$ and the time
coordinate $X_0$ as follows
\begin{equation} \label{action}
S = - \frac{\sqrt{\lambda}}{4 \pi} \int \left[ \frac{1}{2}
\text{tr}(j \wedge \ast j) + dX_0 \wedge \ast dX_0 \right].
\end{equation}
The equations of motion that follow from this action are
\begin{equation} \label{eq of motion}
d \ast j = 0, \quad dj - j \wedge j \equiv 0, \qquad d \ast d X_0 = 0,
\end{equation}
where the second equation is an identity following from the
definition of $j$. To describe physical motions of the string, the
equations of motion \eqref{eq of motion} have to be supplemented
by the Virasoro constraints which in conformal gauge read
\begin{equation} \label{Virasoro no gauge}
\frac{1}{2} \text{tr} j_{\pm}^2 = - (\partial_{\pm} X_0)^2.
\end{equation}
where $j_{\pm} = j_0 \pm j_1$ are the components of the
current $j$ in the worldsheet light-cone coordinates $\sigma^{\pm} =
\frac{1}{2}(\tau \pm \sigma) = \frac{1}{2}(\sigma^0 \pm
\sigma^1)$.
The equation of motion for $X_0$ in \eqref{eq of motion} is
decoupled from the other fields and hence can be solved
separately. In analogy with the flat space case we can then make
use of the residual gauge symmetry to impose say static gauge,
$X_0 = \kappa \tau$, which fixes the $\tau$ coordinate and leaves
only the possibility of rigid translations in the $\sigma$
coordinate
\begin{equation} \label{global sigma}
\sigma \rightarrow \sigma + const.
\end{equation}
Thus, working in conformal static gauge, the original gauge
invariance of the full string action is completely fixed except
for the global transformation \eqref{global sigma} under which
physical states must be invariant. The conserved charges associated
with rigid translations of the worldsheet
coordinate $\sigma$ and $\tau$ are $\mathcal{P}$ and
$\mathcal{E}-\sqrt{\lambda}\kappa^2/2$ where
\begin{equation} \label{defem}
\mathcal{P} = -\frac{\sqrt{\lambda}}{4\pi}\, \int_0^{2\pi}\,
d\sigma\, {\rm tr}\left[ j_0 j_1 \right], \quad
\mathcal{E} = -\frac{\sqrt{\lambda}}{4\pi}\, \int_0^{2\pi}\,
d\sigma\, \frac{1}{2}{\rm tr}\left[ j_0^2 + j_1^2\right].
\end{equation}

In static gauge, since $X_0$ is completely specified, only the
equations of motion for the current $j$ in \eqref{eq of motion}
remain and the Virasoro constraint simplify to
\begin{equation} \label{Virasoro}
\frac{1}{2} \text{tr} j_{\pm}^2 = -\kappa^2.
\end{equation}
It will be convenient to postpone implementing the momentum
constraint $\mathcal{P}=0$ until the very end of the calculation.
Hence we split the constraints \eqref{Virasoro} into two parts.
The first set of constraints read,
\begin{equation} \label{V1}
\frac{1}{2} \text{tr} j_{\pm}^2 = -\kappa_{\pm}^2,
\end{equation}
where $\kappa_{\pm}$ are constants. The world-sheet momentum and
energy \eqref{defem} then become
\begin{equation} \label{EP1}
\mathcal{P} = \frac{\sqrt{\lambda}}{4}(\kappa^2_+ - \kappa^2_-), \quad
\mathcal{E} = \frac{\sqrt{\lambda}}{4}(\kappa^2_+ + \kappa^2_-).
\end{equation}
The remaining content of the Virasoro constraint \eqref{Virasoro} is
the vanishing of the total momentum $\mathcal{P} = 0$ which implies
$\kappa_+^2 = \kappa_-^2 = \kappa^2$ and the string mass-shell
condition,
\begin{equation}\label{massshell}
\mathcal{E}=\frac{\sqrt{\lambda}}{2}\kappa^2 =
\frac{\Delta^2}{2\sqrt{\lambda}}.
\end{equation}

The action \eqref{action} has the following global $SU(2)_L \times
SU(2)_R$ symmetry
\begin{equation*}
g \rightarrow U_L g U_R, \quad U_L,U_R \in SU(2),
\end{equation*}
with corresponding Noether charges
\begin{equation*}
SU(2)_R: \quad Q_R = \frac{\sqrt{\lambda}}{4 \pi} \int_{\gamma}
\ast j, \qquad SU(2)_L: \quad Q_L = \frac{\sqrt{\lambda}}{4 \pi}
\int_{\gamma} \ast \left( g j g^{-1} \right),
\end{equation*}
where $\gamma$ is any closed curve winding once around the
world-sheet. Since these charges are conserved classically,
without loss of generality we can restrict attention to `highest
weight' solutions defined by the level set
\begin{equation} \label{R,L level sets}
Q_R = \frac{1}{2i}R \sigma_3, \; Q_L = \frac{1}{2i}L \sigma_3,
\quad R,L \in \mathbb{R}_+.
\end{equation}
Any other solution with Casimirs $Q_R^2 = R^2, Q_L^2 = L^2$ can be
obtained from a `highest weight' solution by applying to it a
combination of $SU(2)_R$ and $SU(2)_L$ transformations. Note that
the current $j$ is $SU(2)_L$ invariant, but transforms under
$SU(2)_R$ by conjugation
\begin{equation*}
j \rightarrow U_R^{-1} j U_R.
\end{equation*}

\subsection{Hamiltonian framework} \label{section: Hamiltonian}

Starting from the action \eqref{action} we first derive the
Poisson brackets of the system. It will be convenient to choose as
our generalised coordinates, the (target-space) time coordinate
$q^0(\sigma) = X_0(\sigma)$ and the spatial component of the
$SU(2)_R$ current, $q^a(\sigma) = j_1^a(\sigma)$ for $a = 1,2,3$.
To proceed further we first choose a particular basis $t^a$ of the
Lie algebra $\mathfrak{su}(2)$ with structure constants $f^{abc}$
and normalised such that
\begin{equation*}
[t^a,t^b] = f^{abc} t^c, \quad \text{tr}(t^a t^b) = - \delta^{ab}.
\end{equation*}
The action \eqref{action} then reads
\begin{equation*}
S = \frac{\sqrt{\lambda}}{4 \pi} \int d^{\,2} \sigma \left[
\sum_{a = 1}^n \frac{1}{2}\left[ (j_0^a)^2 - (j_1^a)^2 \right] -
\dot{X}^2_0 + X'^2_0 \right].
\end{equation*}
Here the dot and prime denote differentiation with respect to the
worldsheet coordinates $\tau$ and $\sigma$ respectively. The
conjugate momentum for the time coordinate $X_0(\sigma)$ is given
by,
\begin{equation*}
\pi^0(\sigma) = \frac{\delta S}{\delta \dot{q}^0(\sigma)} =
-\frac{\sqrt{\lambda}}{2\pi} \dot{X}_0(\sigma).
\end{equation*}
By the flatness of $j$ it follows that $\dot{q}^a(\sigma) =
\frac{\partial j_0^a}{\partial \sigma} - [j_1,j_0]^a = \nabla_1
j_0^a$, where $\nabla_1$ is the covariant derivative ($\nabla =
d-j$) for the connection $j_1 = j_1^a t^a$. The conjugate momentum
of $q^a(\sigma)$ is then
\begin{align*}
\pi^a(\sigma) &= \frac{\delta S}{\delta \dot{q}^a(\sigma)} = -
\frac{\sqrt{\lambda}}{8 \pi} \int \sum_{b=1}^n \frac{\delta
(j_0^b)^2(\sigma')}{\delta \dot{q}^a (\sigma)} d\sigma' d\tau' = -
\frac{\sqrt{\lambda}}{4 \pi} \int \sum_{b=1}^n \frac{\delta
j_0^b(\sigma')}{\delta \dot{q}^a
(\sigma)} j_0^b(\sigma') d\sigma' d\tau'\\
&= - \frac{\sqrt{\lambda}}{4 \pi} \int \sum_{b=1}^n
\nabla_1^{-1}(\delta_a^b \delta(\sigma - \sigma') \delta(\tau -
\tau')) j_0^b(\sigma') d\sigma' d\tau' = \frac{\sqrt{\lambda}}{4
\pi} \nabla_1^{-1} j_0^a(\sigma).
\end{align*}
In other words, $\frac{\sqrt{\lambda}}{4 \pi} j_0^a(\sigma) =
\nabla_1 \pi^a(\sigma)$ for $a = 1,2,3$. Introducing a new index
$A = 0,1,2,3$, the full set of canonical Poisson brackets between the
generalised coordinates and their conjugate momenta are,
\begin{align*}
&\left\{ q^A(\sigma), q^B(\sigma')\right\} = \left\{ \pi^A(\sigma),
\pi^B(\sigma')\right\} = 0 \\
&\left\{ q^A(\sigma), \pi^B(\sigma')\right\} =
\delta^{AB}\delta(\sigma - \sigma').
\end{align*}
It is convenient to rewrite these Poisson brackets by eliminating
three conjugate momenta $\pi_A$ for $A > 0$, in favour of the
current components $j^a_0$ to obtain,
\begin{subequations} \label{jj Poisson brackets}
\begin{equation} \label{jj PB1}
\left\{ j_1^a(\sigma), j_1^b(\sigma')\right\} = 0,
\end{equation}
\begin{align} \label{jj PB2}
\frac{\sqrt{\lambda}}{4 \pi} \left\{ j_0^a(\sigma),
j_1^b(\sigma')\right\} &= \left\{ \nabla_1
\pi^a(\sigma), q^b(\sigma')\right\} \notag \\
&= - \left\{ q^b(\sigma'), \partial_{\sigma} \pi^a(\sigma) \right\}
+ f^{adc} \left\{ q^b(\sigma'), q^d(\sigma) \pi^c(\sigma)\right\} \notag\\
&= - \partial_{\sigma}\left( \delta^{ab} \delta(\sigma' -
\sigma)\right) + f^{adc} q^d(\sigma) \delta^{bc} \delta(\sigma' - \sigma) \notag\\
 &= - f^{abc} j_1^c(\sigma) \delta(\sigma -
\sigma') - \delta^{ab} \delta'(\sigma - \sigma'),
\end{align}
\begin{equation} \label{jj PB3} \qquad
\frac{\sqrt{\lambda}}{4 \pi} \left\{ j_0^a(\sigma),
j_0^b(\sigma')\right\} = - f^{abc} j_0^c(\sigma) \delta(\sigma -
\sigma').
\end{equation}
\end{subequations}

As expected, the $SU(2)_R$ symmetry is generated by the Noether
charge $Q_R$. Indeed, we find from the last two equations that the
Noether charge $Q_R$ acts on the $SU(2)_R$ current $j$ as follows
\begin{equation} \label{SU(2)_R action}
\left\{ \epsilon \cdot Q_R, j \right\} = \left[ j, \epsilon \right]
= \delta_{\epsilon} j,
\end{equation}
where $\epsilon = \epsilon^a t^a \in \mathfrak{su}(2)$ is
infinitesimal and $\epsilon \cdot Q_R = \epsilon^a Q_R^a$.

In the Hamiltonian formalism, the dynamics of the string is encoded
in the Virasoro constraints,
\begin{equation} \label{Virasoro2}
\begin{split}
\mathcal{H}_{\tau} &= \sum_{a=1}^3 \left[\left(j_0^a\right)^2 +
\left(j_1^b\right)^2 \right] -
\frac{4\pi^2}{\lambda}\left(\pi^{0}\right)^{2} -
X'^2_0 = 0, \\
\mathcal{H}_{\sigma} &= \sum_{a=1}^3 \left(j_0^a j_1^a \right) +
\frac{2 \pi}{\lambda} \pi_0 X'_0 = 0.
\end{split}
\end{equation}
The corresponding Hamiltonian takes the form
\begin{equation*}
H = \frac{1}{2\pi} \int_0^{2\pi} d\sigma \left( \mathcal{N}_{\tau}
\mathcal{H}_{\tau} + \mathcal{N}_{\sigma} \mathcal{H}_{\sigma}
\right),
\end{equation*}
where $\mathcal{N}_{\tau}(\sigma)$ and $\mathcal{N}_{\sigma}(\sigma)$
are Lagrange multipliers for the Virasoro constraints.

As in section \ref{section: strings on S^3} we will choose to work
in static gauge. This corresponds to setting $X_0 = \kappa\tau$
and $\pi_0 = -\sqrt{\lambda}\kappa/2\pi$ where, as before,
$\kappa$ is related to the string energy as $\kappa =
\Delta/\sqrt{\lambda}$. In this gauge, rigid translations of the
world-sheet coordinates $\tau$ and $\sigma$ are generated by the
Hamiltonian functions,
\begin{equation*}
\begin{split}
H^{\rm static}_{\tau} &= \frac{1}{2\pi}\int_0^{2\pi}\, d\sigma\,
\sum_{a=1}^3 \left[\left(j_0^a\right)^2 +
\left(j_1^b\right)^2 \right], \\
H^{\rm static}_{\sigma} &= \frac{1}{2\pi}\int_0^{2\pi}\, d\sigma\,
2\sum_{a=1}^3 \left(j_0^a j_1^a \right).
\end{split}
\end{equation*}
The zero momentum components of the Virasoro constraints
correspond to the string mass-shell condition $H^{\rm
static}_{\tau}=\Delta^{2}/\sqrt{\lambda}$ and the condition that
the total world-sheet momentum should vanish: $H^{\rm
static}_{\sigma}=0$.

Even though the Virasoro constraints are first class by themselves, the
static gauge fixing conditions $X_0 = \kappa\tau$ and $\pi_0 =
-\sqrt{\lambda}\kappa/2\pi$ are second class since $\left\{
X_0(\sigma'), \pi(\sigma) \right\} = \delta(\sigma - \sigma') \neq
0$. This means that to impose all the constraints in the Hamiltonian
framework one must work with Dirac brackets instead of Poisson
brackets\footnote{We are very grateful to Marc Magro for pointing out
this issue.}. However, as we argue in appendix \ref{section: Dirac
brackets}, for the set of action-angle variables which we are
concerned with in this paper the Dirac brackets are the same as their
Poisson brackets since these variables can be defined in a conformally
invariant way.

In the following it will be convenient to think of the
infinite-dimensional phase space of the model, denoted ${\cal P}^{\infty}$
as consisting of all configurations $j_{0}(\sigma),j_{1}(\sigma) \in
\mathfrak{su}(2)$ which obey the Virasoro constraints
\eqref{Virasoro}. Sometimes we will also consider the complexified
phase-space ${\cal P}^{\infty}_{\mathbb{C}}$ with
$j_{0}(\sigma),j_{1}(\sigma) \in \mathfrak{sl}(2,\mathbb{C})$.

\subsection{Conserved charges} \label{section: conserved charges}

The starting point for constructing the infinite tower of
conserved charges for the system is to rewrite its equations of
motion \eqref{eq of motion} as the flatness condition
\begin{equation} \label{zero-curvature}
dJ(x) - J(x) \wedge J(x) = 0,
\end{equation}
for some family of current $J(x)$ on the world-sheet defined in
this case as
\begin{equation*}
J(x) = \frac{1}{1 - x^2} \left( j - x \ast j \right).
\end{equation*}

Owing to the flatness of the current $J(x)$, a natural object to
consider is the parallel transporter on the world-sheet with
$J(x)$ as connection, and in particular the \textit{monodromy
matrix} defined as the parallel transporter around a curve
$c_{\sigma,\tau}$ bound at $(\sigma,\tau)$ and winding once around
the world-sheet
\begin{equation*}
\Omega(x,\sigma,\tau) = P \overleftarrow{\exp}
\int_{[c_{\sigma,\tau}]} J(x),
\end{equation*}
which only depends on the homotopy class $[c_{\sigma,\tau}]$ of
the curve $c_{\sigma,\tau}$ with both end-points fixed at
$(\sigma,\tau)$. An immediate property of $\Omega(x,\sigma,\tau)$
is that its $(\sigma,\tau)$-evolution is isospectral, i.e.
\begin{equation} \label{isospectral}
\Omega(x,\sigma',\tau') = U \Omega(x,\sigma,\tau) U^{-1}, \quad
\text{where } U = P \overleftarrow{\exp}
\int_{(\sigma,\tau)}^{(\sigma',\tau')} J(x).
\end{equation}
This leads straight away to a way of generating infinitely many
conserved charges from traces of powers of monodromy matrices since
\begin{equation*}
\partial_{\sigma} \text{tr }\Omega(x)^n = \partial_{\tau}
\text{tr }\Omega(x)^n = 0.
\end{equation*}

\subsection{Involution of conserved charges}

The statement of the involution property of the conserved charges
generated by $\text{tr }\Omega(x)^n$ is equivalent to the statement
\begin{align*}
\left\{ \text{tr }\Omega(x)^n, \text{tr }\Omega(x')^m \right\} = 0,
\quad \forall n,m \in \mathbb{N}.
\end{align*}
In order to show this we must first obtain the Poisson bracket algebra of
monodromy matrices $\{ \Omega(x) \overset{\otimes}, \Omega(x')
\}$, which is the main focus of this subsection and Appendix
\ref{section: transition matrices PB}. However, as we review below,
since the original
Poisson brackets \eqref{jj PB2} of the current
contain a non-ultralocal term, the resulting
brackets of monodromy matrices are ambiguous and require regularisation.

\subsubsection{Algebra of Lax connections}

The space component of the Lax connection $J(x,\sigma,\tau)$ defined
in section \ref{section: conserved charges} is given by
\begin{equation*}
J_1(\sigma,x) = \frac{1}{1 - x^2} (j_1(\sigma) + x j_0(\sigma)).
\end{equation*}
The monodromy matrix being the path ordered exponential of the
space component $J_1$, we will need the Poisson bracket $\{
J_1,J_1 \}$ in order to construct the Poisson bracket of monodromy
matrices. So consider,
\begin{align*}
&\frac{\sqrt{\lambda}}{4 \pi} \left\{
J_1^a(\sigma,x),J_1^b(\sigma',x')\right\} = -
\frac{\sqrt{\lambda}}{4 \pi} \frac{1}{(1 - x^2)(1 -
{x'}^2)}\left\{ j_1^a(\sigma) + x j_0^a(\sigma), j_1^b(\sigma') +
x' j_0^b(\sigma')
\right\}\\
&= - \frac{1}{(1 - x^2)(1 - {x'}^2)} \left[ (x + x') \left(f^{abc}
j_1^c(\sigma) \delta(\sigma - \sigma') + \delta^{ab} \delta'(\sigma
- \sigma')\right) + xx' f^{abc} j_0^c(\sigma)
\delta(\sigma - \sigma') \right]\\
&= - \frac{x + x'}{(1 - x^2)(1 - {x'}^2)} \delta^{ab} \delta'(\sigma
- \sigma') - \frac{1}{x - x'} \left[ \frac{x^2}{1 - x^2}
J_1^c(\sigma,x') - \frac{{x'}^2}{1 - {x'}^2} J_1^c(\sigma,x) \right]
f^{abc} \delta(\sigma - \sigma').
\end{align*}
Now we switch to tensor notation by contracting both sides with $t^a
\otimes t^b$ and using $[\eta,t^c \otimes \mathbf{1}] = - f^{abc}
t^a \otimes t^b$, $[\eta,\mathbf{1} \otimes t^c] = f^{abc} t^a
\otimes t^b$, where $\eta := - t^a \otimes t^a$, so that
\begin{multline*}
\frac{\sqrt{\lambda}}{4 \pi} \left\{ J_1(\sigma,x)
\mathop{,}^{\otimes} J_1(\sigma',x')\right\} = \left[ -\frac{\eta}{x
- x'}, \frac{{x'}^2}{1 - {x'}^2} J_1(\sigma,x) \otimes {\bf 1} +
\frac{x^2}{1 - x^2} {\bf 1}
\otimes J_1(\sigma,x') \right] \delta(\sigma - \sigma') \\
+ \frac{x + x'}{(1 - x^2)(1 - {x'}^2)} \eta \delta'(\sigma -
\sigma').
\end{multline*}
This bracket has the general form of the fundamental Poisson bracket
$\{ J_1,J_1 \}$ for a non-ultralocal integrable system formulated by
Maillet \cite{Maillet, BFLS}
\begin{multline} \label{Maillet bracket}
\left\{ J_1(\sigma,x) \mathop{,}^{\otimes} J_1(\sigma',x')\right\} =
r'(\sigma,x,x') \delta(\sigma - \sigma')\\ + \left[ r(\sigma,x,x'),
J_1(\sigma,x)\otimes \mathbf{1} + \mathbf{1} \otimes J_1(\sigma',x')
\right] \delta(\sigma - \sigma')\\ - \left[ s(\sigma,x,x'),
J_1(\sigma,x)\otimes \mathbf{1} - \mathbf{1} \otimes J_1(\sigma',x')
\right] \delta(\sigma - \sigma')\\ - \left( s(\sigma,x,x') +
s(\sigma',x,x') \right)\delta'(\sigma - \sigma'),
\end{multline}
where in our case $s(x,x') = - \frac{2 \pi}{\sqrt{\lambda}}
\frac{x+x'}{(1 - x^2)(1 - {x'}^2)} \eta$ is constant (independent of
$\sigma$ and $\tau$), and we find that $r$ is constant as well and
given by
\begin{equation} \label{r-matrix}
r(x,x') = - \frac{2 \pi}{\sqrt{\lambda}} \frac{x^2 + {x'}^2 - 2 x^2
{x'}^2}{(x-x')(1 - x^2)(1 - {x'}^2)} \eta.
\end{equation}
The principal chiral model was first described in terms of Maillet's
$(r-s)$-matrix formalism in \cite{Maillet3}.

\subsubsection{Algebra of monodromy matrices: Maillet regularisation}
\label{section: Maillet}

A first step towards obtaining the algebra of monodromy matrices is to
consider first the algebra of transition matrices. A transition matrix
is defined relative to an interval $[\sigma_1, \sigma_2]$ as follows
\begin{equation*}
T(\sigma_1,\sigma_2,x) = P \overleftarrow{\exp}
\int_{\sigma_2}^{\sigma_1} d\sigma J_1(\sigma,x).
\end{equation*}
The monodromy matrix is then simply a special transition matrix whose
interval wraps the circle fully once, i.e.
\begin{equation*}
\Omega(x,\sigma) = T(\sigma,\sigma + 2 \pi,x).
\end{equation*}
Since the derivation of the algebra of transition matrices is fairly
standard \cite{Faddeev:1987ph} we have left it to Appendix
\ref{section: transition matrices PB} to avoid cluttering this section
with algebra. The end result is the following bracket between two
transition matrices with distinct intervals,
\begin{multline*}
\left\{ T(\sigma_1,\sigma_2,x) \mathop{,}^{\otimes}
T(\sigma'_1,\sigma'_2,x') \right\} \\
\begin{split}
= &+ \epsilon(\sigma'_1 - \sigma'_2) \chi(\sigma;
\sigma'_1,\sigma'_2)
\\ &\times \left. T(\sigma_1,\sigma,x) \otimes
T(\sigma'_1,\sigma,x') \left( r(\sigma,x,x') - s(\sigma,x,x')
\right) T(\sigma,\sigma_2,x) \otimes T(\sigma,\sigma'_2,x')
\right|_{\sigma = \sigma_2}^{\sigma =
\sigma_1} \\
&+ \epsilon(\sigma_1 - \sigma_2) \chi(\sigma; \sigma_1,\sigma_2)
\\ &\times \left. T(\sigma_1,\sigma,x) \otimes T(\sigma'_1,\sigma,x')
\left( r(\sigma,x,x') + s(\sigma,x,x') \right) T(\sigma,\sigma_2,x)
\otimes T(\sigma,\sigma'_2,x') \right|_{\sigma = \sigma'_2}^{\sigma
= \sigma'_1}.
\end{split}
\end{multline*}
It follows from this algebra that the function,
\begin{equation*}
\Delta^{(1)}(\sigma_1,\sigma_2,\sigma'_1,\sigma'_2; x,x') = \{
T(\sigma_1,\sigma_2,x) \, \overset{\otimes}, \,
T(\sigma'_1,\sigma'_2,x') \}
\end{equation*}
is well defined and continuous where $\sigma_1, \sigma_2, \sigma'_1,
\sigma'_2$ are all distinct, but it has discontinuities proportional
to $2 s$ precisely across the hyperplanes corresponding to some of
the $\sigma_1, \sigma_2, \sigma'_1, \sigma'_2$ being equal. Defining
the Poisson bracket $\{ T \, \overset{\otimes}, \,  T\}$ for
coinciding intervals ($\sigma_1 = \sigma'_1, \sigma_2 = \sigma'_2$)
or adjacent intervals ($\sigma'_1 = \sigma_2$ or $\sigma_1 =
\sigma'_2$) requires defining the value of the discontinuous
matrix-valued function $\Delta^{(1)}$ at its discontinuities. It is
shown in \cite{Maillet} that requiring antisymmetry of the Poisson
bracket and the derivation rule to hold imposes the symmetric
definition of $\Delta^{(1)}$ at its discontinuous points; for
example at $\sigma_1 = \sigma'_1$ we must define
\begin{multline*}
\Delta^{(1)}(\sigma_1,\sigma_2,\sigma_1,\sigma'_2; x,x') \\ =
\lim_{\epsilon \rightarrow 0^+} \frac{1}{2} \left(
\Delta^{(1)}(\sigma_1,\sigma_2,\sigma_1 + \epsilon,\sigma'_2; x,x')
+ \Delta^{(1)}(\sigma_1,\sigma_2,\sigma_1 - \epsilon,\sigma'_2;
x,x') \right),
\end{multline*}
and likewise for all other possible coinciding endpoints. This
definition is equivalent to assigning the value of $\frac{1}{2}$ to
the characteristic function $\chi$ at its discontinuities. Having
thus defined $\Delta^{(1)}$ at its discontinuities we now have a
definition of the Poisson bracket $\{ T \, \overset{\otimes}, \,
T\}$ for coinciding and adjacent intervals consistent with the
antisymmetry of the Poisson bracket and the derivation rule. However
this definition of the $\{ T \, \overset{\otimes}, \, T\}$ Poisson
bracket does not satisfy the Jacobi identity as is shown in
\cite{Maillet}, so that in fact no strong definition of the bracket
$\{ T \, \overset{\otimes}, \, T\}$ with coinciding or adjacent
intervals can be given without violating the Jacobi identity
\cite{Maillet}. It is nevertheless possible \cite{Maillet, Maillet2}
to give a weak\footnote{The bracket is weak
in the sense that any multiple Poisson
bracket of $T$'s can be given a meaning which cannot be reduced to
its similarly defined constituent Poisson brackets, i.e. the
multiple Poisson bracket $\{ T \, \overset{\otimes}, \, \{ \ldots \{
T \, \overset{\otimes}, \, T \} \ldots \} \}$ with $n$ factors of
$T$ must be separately defined for each $n$.} definition of this
bracket for coinciding or adjacent intervals in a way that is
consistent with the Jacobi identity as follows: consider the
multiple Poisson bracket of $(n+1)$ transition matrices
\begin{multline*}
\Delta^{(n)}\left(\sigma^{(1)}_1, \sigma^{(1)}_2, \ldots,
\sigma^{(n+1)}_1, \sigma^{(n+1)}_2; x^{(1)}, \ldots,x^{(n+1)}\right) \\
= \left\{ T\left(\sigma^{(1)}_1,\sigma^{(1)}_2,x^{(1)}\right) \,
\overset{\otimes}, \, \left\{ \ldots \, \overset{\otimes}, \,
\left\{ T\left(\sigma^{(n)}_1,\sigma^{(n)}_2,x^{(n)}\right) \,
\overset{\otimes}, \,
T\left(\sigma^{(n+1)}_1,\sigma^{(n+1)}_2,x^{(n+1)}\right) \right\}
\ldots \right\} \right\},
\end{multline*}
which is unambiguously defined and continuous where $\sigma^{(1)}_1,
\sigma^{(1)}_2, \ldots, \sigma^{(n+1)}_1, \sigma^{(n+1)}_2$ are all
distinct, but again is discontinuous across the hyperplanes defined
by some of the points $\sigma^{(1)}_1, \sigma^{(1)}_2, \ldots,
\sigma^{(n+1)}_1, \sigma^{(n+1)}_2$ being equal. The values of
$\Delta^{(n)}$ at its discontinuities are defined by employing a
point splitting regularisation followed by a total symmetrisation
limit \cite{Maillet}. For example, we define its value at
$\sigma^{(i)}_1 = \sigma_1, i = 1,\ldots,n+1$ by
\begin{multline*}
\Delta^{(n)}\left(\sigma_1, \sigma^{(1)}_2, \ldots, \sigma_1,
\sigma^{(n+1)}_2; x^{(1)}, \ldots,x^{(n+1)}\right) \\ =
\lim_{\epsilon \rightarrow 0^+} \frac{1}{(n+1)!} \sum_{p \in
S_{n+1}} \Delta^{(n)}\left(\sigma_1 + p(1) \epsilon, \sigma^{(1)}_2,
\ldots, \sigma_1 + p(n+1) \epsilon, \sigma^{(n+1)}_2; x^{(1)},
\ldots,x^{(n+1)}\right),
\end{multline*}
and similarly one defines the value of $\Delta^{(n)}$ at all other
discontinuities. With the function $\Delta^{(n)}$ being defined at
its discontinuities we now have the definition of a weak bracket
which reduces to the normal Poisson bracket on quantities for which
the latter is continuous. It is shown in \cite{Maillet} that the
Jacobi identity for transition matrices with coinciding or adjacent
interval is now satisfied in terms of this weak bracket
($\Delta^{(2)}$ being the relevant quantity in this case).

\begin{figure}
\centering \psfrag{g}{\footnotesize $\gamma$}
\psfrag{gp}{\footnotesize $\gamma'$} \psfrag{S1}{\footnotesize
$S^1$} \psfrag{g2}{\footnotesize $\tilde{\gamma}$}
\psfrag{gp2}{\footnotesize $\tilde{\gamma}'$}
\psfrag{gp3}{\footnotesize $\tilde{\gamma}'$}
\psfrag{0}{\footnotesize $0$} \psfrag{2p}{\footnotesize$2 \pi$}
\psfrag{R}{\footnotesize $\mathbb{R}$}
\includegraphics[height=40mm,width=110mm]{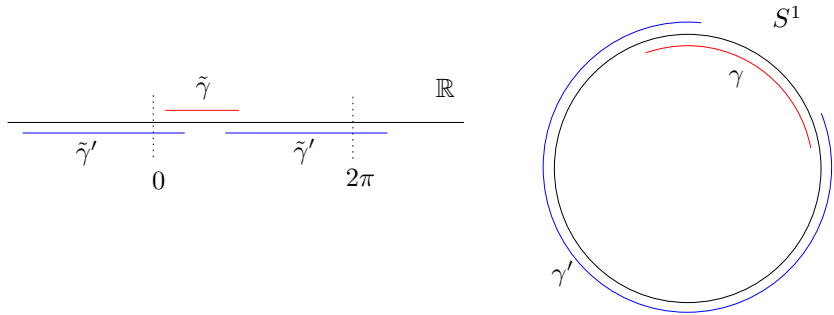}
\caption{Example of a path lifting required in computing Poisson
brackets of transition matrices on $S^1$ of the form $\{ T(\gamma,x)
\, \overset{\otimes}, \, T(\gamma',x') \}$.} \label{circle PB}
\end{figure}

Using this regularisation procedure we now derive an expression for
the Poisson bracket between two monodromy matrices in the periodic
case under consideration, a result which was first obtained in
\cite{Maillet2, Maillet}. To begin with consider the Poisson bracket
$\{ T(\gamma,x) \, \overset{\otimes}, \, T(\gamma',x') \}$ between two
generic transition matrices $T(\gamma,x)$ and $T(\gamma',x')$ on the
circle $S^1$, defined relative to two different paths $\gamma$ and
$\gamma'$ on $S^1$, e.g.
\begin{equation} \label{generic transition matrix}
T(\gamma,x) = P \overleftarrow{\exp} \int_{\gamma} d\sigma
J_1(\sigma,x).
\end{equation}
We would like to compute this bracket by working on the universal
cover $\mathbb{R}$ of $S^1$. So we choose a lift $\tilde{\gamma}$ of
the path $\gamma$ to $\mathbb{R}$. Then because the only
contribution to the Poisson bracket comes from the region of overlap
between $\gamma$ and $\gamma'$ on $S^1$ (by \eqref{PB definition}
and \eqref{generic transition matrix}), we have that
\begin{equation} \label{PB on S^1}
\{ T(\gamma,x) \, \overset{\otimes}, \, T(\gamma',x') \} =
\sum_{\tilde{\gamma}' \text{ lift of } \gamma'} \{
T(\tilde{\gamma},x) \, \overset{\otimes}, \, T(\tilde{\gamma}',x')
\},
\end{equation}
where the sum is over lifts $\tilde{\gamma}'$ of $\gamma'$ to
$\mathbb{R}$. An example of these lifted paths is shown in Figure
\ref{circle PB}.
Let us now apply this formula to compute the Poisson bracket between
two transition matrices $\Omega(x,\sigma)$ and
$\Omega(x',\sigma)$ on $S^1$. The common interval $\gamma$ of
both matrices stretches once around the full circle and so it
follows that if we take $\tilde{\gamma} = [\sigma,\sigma + 2\pi]$ to
be the lift of the interval of $\Omega(x,\sigma)$ then there
are only three possibilities for the lift $\tilde{\gamma}'$ of the
interval of $\Omega(x',\sigma)$ which give a non-zero
contribution to the right hand side of \eqref{PB on S^1}, namely
\begin{equation} \label{contributing intervals}
[\sigma - 2 \pi,\sigma], \quad [\sigma,\sigma + 2 \pi], \quad
[\sigma + 2 \pi,\sigma + 4 \pi].
\end{equation}
Since the corresponding three brackets $\{ T(\tilde{\gamma},x) \,
\overset{\otimes}, \, T(\tilde{\gamma}',x') \}$ on $\mathbb{R}$
are over coinciding or adjacent intervals they need to be
regularised by the procedure described above. Let us start by
considering the coinciding interval bracket $\{ T(\sigma,\sigma +
2 \pi,x) \, \overset{\otimes}, \, T(\sigma,\sigma + 2 \pi,x') \}$.
There are 4 different possible point splittings of the endpoints,
each giving the same contribution (using \eqref{transition matrix
PB})
\begin{equation*}
r(x,x') \left( \Omega(x,\sigma) \otimes \Omega(x',\sigma)
\right) - \left( \Omega(x,\sigma) \otimes
\Omega(x',\sigma) \right) r(x,x')
\end{equation*}
in the limit of coinciding points. On the other hand, the adjacent
interval brackets (corresponding to the first and last choices for
$\tilde{\gamma}'$ in \eqref{contributing intervals}) each have two
possible point splittings and together they contribute, in the
coinciding end-point limit,
\begin{equation*}
\left( \Omega(x,\sigma) \otimes {\bf 1} \right) s(x,x') \left(
{\bf 1} \otimes \Omega(x',\sigma) \right) - \left( {\bf 1}
\otimes \Omega(x',\sigma) \right) s(x,x') \left(
\Omega(x,\sigma) \otimes {\bf 1} \right)
\end{equation*}
to the Poisson bracket of two monodromy matrices. The sum of the
last two expressions gives the right hand side of \eqref{PB on S^1}
which yields the sought-after (weak) Poisson bracket between two
monodromy matrices on $S^1$
\begin{align} \label{fundamental Poisson bracket}
\left\{ \Omega(x,\sigma) \mathop{,}^{\otimes}
\Omega(x',\sigma) \right\} = &[r(x,x'), \Omega(x,\sigma)
\otimes \Omega(x',\sigma)] \notag \\ +
&\left(\Omega(x,\sigma) \otimes {\bf 1}\right) s(x,x') \left(
{\bf 1} \otimes \Omega(x',\sigma) \right) \notag
\\ - &\left( {\bf 1} \otimes \Omega(x',\sigma) \right) s(x,x') \left(
\Omega(x,\sigma) \otimes {\bf 1} \right).
\end{align}
Consider now the bracket $\left\{ \Omega(x,\sigma)^n
\mathop{,}^{\otimes} \Omega(x',\sigma)^m \right\}$ for any $n,m
\in \mathbb{N}$, which can easily be reduced to the previous Poisson
bracket as follows (omitting the $\sigma$-dependence)
\begin{equation*}
\left\{ \Omega(x)^n \mathop{,}^{\otimes}
\Omega(x')^m \right\} = nm \left( \Omega(x)^{n-1}
\otimes {\bf 1} \right) \left\{ \Omega(x)
\mathop{,}^{\otimes} \Omega(x') \right\} \left( {\bf 1}
\otimes \Omega(x)^{m-1} \right).
\end{equation*}
Then using the standard notational shorthands
$\overset{1}A = A \otimes {\bf 1}$ and $\overset{2}A = {\bf 1} \otimes
A$, and taking the trace over both factors of the tensor product we find
\begin{align*}
\left\{ \text{tr } \Omega(x)^n, \text{tr } \Omega(x')^m \right\} &=
nm \text{ tr}_{12} \left( \overset{1}\Omega(x)^{n-1}
\overset{2}\Omega(x')^{m-1} \left\{ \overset{1} \Omega(x),
\overset{2}\Omega(x') \right\} \right) \\
&= nm \text{ tr}_{12} \left[r(x,x') + s(x,x'),
\overset{1}\Omega(x)^n \overset{2}\Omega(x')^m \right] \\
\text{i.e. } \left\{ \text{tr } \Omega(x)^n, \text{tr } \Omega(x')^m
\right\} &= 0,
\end{align*}
where in the second line we have used \eqref{fundamental Poisson
bracket}. Because this last bracket is zero it can be
understood as defining a bracket in the strong sense and without
recourse to any regularisation. We deduce from this last relation
that the invariants of the system encoded in the quantity $\text{tr
} \Omega(x)^n$ are in involution with respect to the Poisson
bracket; this is the full statement of Liouville integrability of
the system.

As a specific check of \eqref{fundamental Poisson bracket} we show
that the $SU(2)_R$ symmetry is canonically realised on $\Omega(x)$
via the weak Poisson bracket \cite{Maillet2}. It is
straightforward to show that the monodromy matrix has the
following asymptotics at $x = \infty$ \cite{Paper1},
\begin{equation*}
\Omega(x,\sigma,\tau) = {\bf 1} + \frac{1}{x} \frac{4 \pi
Q_R}{\sqrt{\lambda}} + O\left( \frac{1}{x^2} \right) \quad \text{as
} x \rightarrow \infty.
\end{equation*}
Then starting with equation \eqref{fundamental Poisson bracket}
multiplied by $x \, (\epsilon \otimes {\bf 1})$ and taking the
trace over the first tensor product space followed by the limits
$x \rightarrow \infty$ and $x' \rightarrow 0$ one deduces, using
also the asymptotics $r(x,x') \sim_{x \rightarrow \infty} \frac{2
\pi}{\sqrt{\lambda}} \frac{1-2 x'^2}{x(1-x'^2)}$ and $s(x,x')
\sim_{x \rightarrow \infty} \frac{2 \pi}{\sqrt{\lambda}}
\frac{1}{x(1-x'^2)}$, that
\begin{equation*}
\left\{ \epsilon \cdot Q_R , \Omega(x') \right\} = \left[ \epsilon,
\Omega(x') \right].
\end{equation*}
In other words, the right Noether charge $Q_R$ generates the
correct transformation on $\Omega(x)$, which we expect to be
\begin{equation*}
\Omega(x) \rightarrow U_R^{-1} \Omega(x) U_R,
\end{equation*}
provided we use the weak bracket instead of the Poisson bracket.

\section{Symplectic structure for finite-gap solutions}
\label{section: data}

In a previous paper \cite{Paper1} we constructed the general
finite-gap solution to the equations of motion of a string moving
on $\mathbb{R} \times S^3$ satisfying the Virasoro constraints.
We also constructed the corresponding moduli space of solutions.
Our aim here is to determine the symplectic structure induced on the
moduli space of solutions by the regularised Poisson brackets obtained
in the previous section. As we will see below, our analysis for strings
moving on ${\mathbb{R}}\times S^{3}$ can be thought of as a non-linear
generalisation of the more familiar Hamiltonian analysis of strings in
flat space. We will therefore begin by reviewing the standard
discussion of the flat space case following eg Chapter 2 of
\cite{GSWI}.
\paragraph{}
We will consider a closed bosonic string moving on $(D+1)$-dimensional
Minkowski space with worldsheet fields $X^{\mu}(\sigma,\tau)$ for
$\mu=0,1,\ldots,D$. In conformal gauge, the worldsheet equation of
motion is the two-dimensional Laplace equation $\partial_+ \partial_-
X^{\mu} = 0$. As the equation is linear, the general
solution for closed string boundary conditions is given by the
Fourier series,
\begin{equation} \label{General solution: flat}
X^{\mu}(\sigma,\tau) = x^{\mu} + p^{\mu} \tau + i \sum_{n \neq 0}
\frac{1}{n} \alpha_n^{\mu} e^{-i n(\tau - \sigma)} + i \sum_{n
\neq 0} \frac{1}{n} \tilde{\alpha}_n^{\mu} e^{-i n(\tau +
\sigma)}.
\end{equation}
where the Fourier coefficients $\alpha^{\mu}_{n}$ and
$\tilde{\alpha}^{\mu}_{n}$  correspond to classical oscillator coordinates
for left- and right-moving modes respectively.
For our purposes it will be convenient to restrict our attention to
classical solutions with a finite number of oscillators turned on.
Generic solutions can then be obtained as a limiting case.
We will see that these `finite-oscillator' solutions are close
analogs of the finite-gap solutions of string theory on
$\mathbb{R}\times S^{3}$ and other classically integrable backgrounds.
\paragraph{}
Since \eqref{General solution: flat} is the general solution to
the field equations, the fields
$X^{\mu}(\sigma)=X^{\mu}(\sigma,0)$ and
$P^{\mu}(\sigma)=\dot{X}^{\mu}(\sigma,0)$ restricted to a
$\tau$-slice (taken at $\tau = 0$ without loss of generality) give
a convenient parametrisation of the phase space of the
string\footnote{Note that this is not the physical phase space as
we have not yet imposed the Virasoro constraints.}. In terms of
the oscillator coordinates we find,
\begin{equation}
\label{generic PS configuration}
\begin{split}
X^{\mu}(\sigma) &= x^{\mu} + i \sum_{n \neq 0} \frac{1}{n}
\alpha_n^{\mu} e^{i n\sigma} + i \sum_{n \neq 0}
\frac{1}{n} \tilde{\alpha}_n^{\mu} e^{-i n\sigma},\\
P^{\mu}(\sigma) &= p^{\mu} + \sum_{n \neq 0} \alpha_n^{\mu} e^{i
n\sigma} + \sum_{n \neq 0} \tilde{\alpha}_n^{\mu} e^{-i n\sigma}.
\end{split}
\end{equation}
Conversely the oscillator
coefficients $\alpha_n^{\mu}, \tilde{\alpha}_n^{\mu}$ as well
as the centre of mass position and momenta $x^{\mu}, p^{\mu}$ can
be extracted from a generic phase-space configuration
$X^{\mu}(\sigma), P^{\mu}(\sigma)$ by the following relations
\begin{equation} \label{extract data: flat}
\left\{
\begin{split}
&\alpha_m^{\mu} = \frac{1}{2 \pi} \int_0^{2 \pi} e^{- i m \sigma}
\frac{1}{2} \left( P^{\mu}(\sigma) - \partial_{\sigma}
X^{\mu}(\sigma)
\right) d\sigma, \quad m \neq 0 \\
&\tilde{\alpha}_m^{\mu} = \frac{1}{2 \pi} \int_0^{2 \pi} e^{i m
\sigma} \frac{1}{2} \left( P^{\mu}(\sigma) + \partial_{\sigma}
X^{\mu}(\sigma) \right) d\sigma, \quad m \neq 0 \\
&x^{\mu} = \frac{1}{2 \pi} \int_0^{2 \pi} X^{\mu}(\sigma) d\sigma,
\qquad p^{\mu} = \frac{1}{2 \pi} \int_0^{2 \pi} P^{\mu}(\sigma)
d\sigma.
\end{split}
\right.
\end{equation}
\paragraph{}
Equations
\eqref{extract data: flat} are the inverse of the equations
\eqref{generic PS configuration} and the transformation
\begin{equation} \label{flat space data}
\left\{ X^{\mu}(\sigma), P^{\mu}(\sigma) \right\} \mapsto
\left\{ x^{\mu}, p^{\mu}, \alpha_n^{\mu}, \tilde{\alpha}_n^{\mu}
\right\}
\end{equation}
is simply a change of variable on phase-space.
The Poisson brackets which follow from the string action, take the
form,
\begin{gather} \label{canonical PB: flat}
\{ X^{\mu}(\sigma), X^{\nu}(\sigma') \} = \{
P^{\mu}(\sigma),
P^{\nu}(\sigma') \} = 0,\\
\{ P^{\mu}(\sigma), X^{\nu}(\sigma') \} = \eta^{\mu
\nu} \delta(\sigma - \sigma'),
\end{gather}
and it is straightforward to rewrite these brackets in the new
coordinate system as,
\begin{equation} \label{data PB: flat}
\begin{split}
\{ \alpha^{\mu}_m, \alpha^{\nu}_n \} &=
i m \delta_{m+n} \eta^{\mu \nu}, \{ \alpha^{\mu}_m,
\tilde{\alpha}^{\nu}_n \} = 0,\\
\{ \tilde{\alpha}^{\mu}_m, \tilde{\alpha}^{\nu}_n
\} &= i m \delta_{m+n} \eta^{\mu \nu}, \{ p^{\mu}, x^{\nu} \} = \eta^{\mu \nu}.
\end{split}
\end{equation}
\paragraph{}
So far we have discussed the full solution space of the equations
of motion. The next step is to restrict to physical configurations
of the string by imposing the Virasoro constraints and fixing the
residual gauge symmetry. These steps are easily accomplished by
imposing light-cone gauge. We begin by defining light-cone
coordinate $X^{\pm}=X^{0}\pm X^{D}$ and imposing the gauge
condition $X^{+}=p^{+}\tau+ x^{+}$. With this choice, it is
possible to solve the Virasoro constraint explicitly to eliminate
the longitudinal oscillator coordinates. The remaining independent
degrees of freedom are,
\begin{equation}
\{x^{i},p^{i},x^{-},p^{-},\alpha^{i}_{n}, \tilde{\alpha}^{i}_{n}\}
\end{equation}
where the index $i=1,2,\ldots,D-1$ runs over the transverse
spacetime dimensions. To find the Poisson brackets of the physical
degrees of freedom one must follow the standard Dirac procedure
for constrained systems. In the present case this is described in
detail in the book by Brink and Henneaux \cite{BrH}. The Virasoro
constraint and the light cone gauge fixing condition together
correspond to a system of second order constraints on the phase
space. Fortunately, the resulting Dirac bracket for the transverse
degrees of freedom is the same as their naive Poisson bracket,
\begin{equation}
\begin{split}
\{ \alpha^{i}_m, \alpha^{j}_n \} &=
i m \delta_{m+n} \delta^{ij}, \{ \alpha^{i}_m,
\tilde{\alpha}^{j}_n \} = 0,\\
\{ \tilde{\alpha}^{i}_m, \tilde{\alpha}^{j}_n
\} &= i m \delta_{m+n} \delta^{i j}, \{ p^{i}, x^{j} \} = \delta^{i j}.
\end{split}
\end{equation}
These Poisson brackets are the starting point for canonical
quantisation of the string which proceeds by the usual recipe of
promoting Poisson brackets to commutators.
\paragraph{}
Classical string theory in flat space is trivially integrable as
the corresponding equation of motion is linear. For comparison
with the non-linear case, it will be convenient to exhibit
integrability explicitly by constructing the corresponding
action-angle variables. As the dynamics of the COM degrees of
freedom of the string is free we will focus on the transverse
oscillators which describe the physical excitations of the string
in its rest frame. We introduce a new set of variables $\{\theta
^j_n, S^j_n, \tilde{\theta}^j_n, \tilde{S}^j_n \}$ for
$j=1,\ldots, D-1$ via the relations,
\begin{equation*}
\alpha_n^j = \sqrt{n S_n^j} e^{i \theta_n^j}, \tilde{\alpha}_n^j =
\sqrt{n \tilde{S}_n^j} e^{i \tilde{\theta}_n^j}.
\end{equation*}
The variables $S^j_n$ and $\tilde{S}^j_n$ are classical analogs of
the occupation numbers of each transverse oscillator. These
variables are trivially time independent and therefore correspond
to conserved charges. One may also check the involution condition,
\begin{equation}
\{S^i_n,S^j_m\}=
\{S^i_n,\tilde{S}^j_m\}=\{\tilde{S}^i_n,\tilde{S}^j_m\}=0
\end{equation}
These are the action variables of the flat space string.
\paragraph{}
The angular variables $\theta^j_n$ and $\tilde{\theta}^j_n$ each
have period $2\pi$ and are canonically conjugate to the
corresponding action variables. Their non-vanishing Poisson
brackets are,
\begin{equation}
\{S^i_n,\theta^j_m\}=\{\tilde{S}^i_n,\tilde{\theta}^j_m\}=
\delta_{nm}\delta^{ij}
\end{equation}
It follows immediately from Hamilton's equations that the angle
variables evolve linearly in time while, as above, the conjugate
action variables remain constant.
\begin{gather*}
\theta_n^{\mu}(\tau) = \theta_n^{\mu}(0) - n \tau, \quad
S_n^{\mu}(\tau)
= S_n^{\mu}(0) = \text{const.}\\
\tilde{\theta}_n^{\mu}(\tau) = \tilde{\theta}_n^{\mu}(0) - n \tau,
\quad \tilde{S}_n^{\mu}(\tau) = \tilde{S}_n^{\mu}(0) =
\text{const.}
\end{gather*}
\paragraph{}
We now turn to the case at hand of string on $\mathbb{R} \times
S^3$ and present a non-linear analogue of the above concepts for
strings on flat space. However we will proceed in a slightly different
order. As we have already fixed the gauge completely in section
\ref{section: strings on S^3} by imposing static gauge $X_0 = \kappa
\tau$, we now solve both the equations of motion and the
Virasoro constraint simultaneously through algebro-geometric methods
in section \ref{section: finite-gap}. In this way we immediately obtain
the general `finite-gap' solution (analogue of the `finite-oscillator'
solution above) expressed directly in terms of physical degrees of
freedom. Section \ref{section:
extract data} aims to derive the analogues of \eqref{extract data:
flat} in the case of the nonlinear differential equations for a
string on $\mathbb{R} \times S^3$ which will be crucial in section
\ref{section: PB data 2} for determining Poisson brackets on the
algebro-geometric data.  Finally, in section \ref{section:
action-angle} we define the change of variable to action-angle
coordinates and verify the canonical Poisson brackets for these
variables.

\subsection{Finite-gap integration} \label{section: finite-gap}

In this subsection we will briefly review the explicit
construction of finite-gap solutions given in \cite{Paper1}. The
reader should consult this reference for extra details.

\subsubsection{The spectral curve} \label{section: spectral curve}

The starting point for the method of finite-gap integration
is the formulation of the equations of motion \eqref{eq of motion}
of the system as the flatness condition \eqref{zero-curvature}.
Representing the equations of motion in this form introduces a large
amount of spurious symmetries which we are free to fix as we proceed;
indeed, equation \eqref{zero-curvature} is invariant under gauge
transformations $J(x) \mapsto \tilde{g} J(x) \tilde{g}^{-1} +
d\tilde{g} \tilde{g}^{-1}$.

Now the isospectral $(\sigma,\tau)$-evolution \eqref{isospectral} of
the monodromy matrix leads naturally to the definition of a
$(\sigma,\tau)$-independent \textit{spectral curve} in $\mathbb{C}^2$
\begin{equation*}
\Gamma : \quad \Gamma(x,y) = \text{det}\left( y {\bf 1} -
\Omega(x,\sigma,\tau) \right) = 0.
\end{equation*}
However, this curve is highly singular \cite{Paper1} and so one
should replace it with an algebraic curve $\Sigma$ defined as a
desingularisation of $\Gamma$ (for details of this construction
see \cite{Paper1,Beisert:2004ag}). An important property of
the spectral curve $\Gamma$ is that above any non-singular point
$(x,y) \in \Gamma$ ($d\Gamma(x,y) \neq 0$) there is a unique
eigenvector of $\Omega(x)$ with eigenvalue $y$. It follows that the
desingularised curve $\Sigma$ has a unique eigenvector above any of
its points. In the present case where $\Omega(x)$ is $2 \times 2$ the
curve $\Sigma$ is also hyperelliptic with projection denoted
$\hat{\pi} : \Sigma \rightarrow \mathbb{CP}^1$; we also introduce the
notation $\{ x^{\pm} \} = \hat{\pi}^{-1}(x)$ for the set of points
above $x \in \mathbb{CP}^1$.

The curve $\Sigma$ is naturally equipped with a normalised second
kind Abelian differential $dp$, with singularities only at the
points $\{ (+1)^{\pm}, (-1)^{\pm} \}$ above $x = \pm 1$, specified
uniquely by
\begin{equation*}
\int_{a_i} dp = 0, \quad \int_{b_i} dp = 2 \pi k_i \in \mathbb{Z},
\end{equation*}
\begin{equation} \label{dp asymp at pm 1}
\begin{split}
dp(x^{\pm}) &= \mp d\left(\frac{\pi \kappa_+}{x - 1}\right) +
O\left( (x -
1)^0 \right) \quad \text{as } x \rightarrow +1, \\
dp(x^{\pm}) &= \mp d\left(\frac{\pi \kappa_-}{x + 1}\right) +
O\left( (x + 1)^0 \right) \quad \text{as } x \rightarrow -1,
\end{split}
\end{equation}
where $\{ a_i, b_i\}_{i=1}^g$ is a canonical basis of
$H_1(\Sigma,\mathbb{Z})$. The asymptotics of $dp$ near the points
$\{ 0^{\pm}, \infty^{\pm} \}$ can be deduced \cite{Paper1} from
the `highest weight' conditions \eqref{R,L level sets} and are
directly related to the Casimirs $R^2, L^2$ of $SU(2)_R \times
SU(2)_L$
\begin{equation} \label{dp asymp at 0,infty}
\begin{split}
dp(x^{\pm}) &= \mp d \left[\frac{1}{x} \frac{2 \pi
R}{\sqrt{\lambda}} + O\left(
\frac{1}{x^2} \right)\right], \quad \text{as } x \rightarrow \infty, \\
dp(x^{\pm}) &= \pm d \left[x \frac{2 \pi L}{\sqrt{\lambda}} +
O\left( x^2 \right)\right], \quad \text{as } x \rightarrow 0.
\end{split}
\end{equation}
The Abelian integral $p(P) = \int_{\infty^+}^P dp$ is called the
quasi-momentum and has the property that $\{ e^{i p(x^+)}, e^{i
p(x^-)} \}$ are the eigenvalues of $\Omega(x)$.

A convenient way of describing the moduli space of genus $g$
curves $\Sigma$ with punctures at $\{ 1^{\pm}, (-1)^{\pm},
0^{\pm}, \infty^{\pm} \}$ and equipped with a meromorphic
differential $dp$ with specified behaviours \eqref{dp asymp at pm
1}, \eqref{dp asymp at 0,infty} near these punctures is as a leaf
$\mathcal{L}$ in a foliation of the universal configuration
space of \cite{Krichever+Phong}. To make contact with the
construction of the universal configuration space of
\cite{Krichever+Phong} we also introduce another meromorphic
differential $dz$ by specifying its Abelian integral
\begin{equation*}
z = x + \frac{1}{x},
\end{equation*}
which is a single-valued function on $\Sigma$ so that all periods
of $dz$ are zero (i.e. $\int_C dz = 0$ for any cycle $C \in
H_1(\Sigma,\mathbb{R})$). The asymptotics of $dz$ near the
punctures are obvious from its definition. Full details of this
construction can be found in \cite{Paper1}. However, in
\cite{Paper1} we chose to keep $R$ fixed, thereby describing only
the internal degrees of freedom of the string by a leaf
$\mathcal{L}|_R$ in a smooth $g$-dimensional foliation of the
universal moduli space; in the present paper we allow $R$ to vary
so the leaf $\mathcal{L}$ under consideration will now have one
extra dimension.

Using the set of local coordinates on this universal configuration
space introduced in \cite{Krichever+Phong} the leaf in question is
obtained as the joint level set of all but $g + 1$ of the
coordinates. Defining the following differential on $\Sigma$
\begin{equation} \label{symplectic 1-form}
\quad \alpha = \frac{\sqrt{\lambda}}{4 \pi} z dp,
\end{equation}
the remaining $g+1$ coordinates parametrising the leaf are
\cite{Krichever+Phong}
\begin{equation} \label{independent moduli}
S_i = \frac{1}{2 \pi i} \int_{a_i} \alpha, \; i = 1, \ldots, g,
\quad \frac{R}{2} = - \text{Res}_{\infty^{+}} \alpha.
\end{equation}
Equivalently one can parametrise the moduli space $\mathcal{L}$ by
assigning a \textit{filling fractions} to each of the $K = g + 1$
cuts $\mathcal{C}_I$
\begin{equation} \label{filling fractions}
\mathcal{S}_I = \frac{1}{2 \pi i} \int_{\mathcal{A}_I} \alpha, \;
I = 1, \ldots, K = g+1,
\end{equation}
where $\mathcal{A}_I$ is a cycle encircling the cut
$\mathcal{C}_I$ on the physical sheet. The filling fractions are
related to the variable $R$ and the parameter $\frac{L}{2} =
\text{Res}_{0^{+}} \alpha$ by
\begin{equation*}
\sum_{I=1}^K \mathcal{S}_I = \text{Res}_{\infty^{+}} \alpha +
\text{Res}_{0^{+}} \alpha = \frac{1}{2} (L - R).
\end{equation*}
The moduli space $\mathcal{L}$ is therefore a complex manifold
with only orbifold singularities of dimension
\begin{equation*}
\text{dim } \mathcal{L} = g+1,
\end{equation*}
every point of which corresponds to an admissible pair
$(\Sigma,dp)$ where $\Sigma$ has genus $g$.

\subsubsection{The normalised eigenvector} \label{section: The normalised eigenvector}

Let us denote by $\bm{h}(P,\sigma,\tau)$ the unique normalised
eigenvector of $\Omega(x,\sigma,\tau)$ at a point $P \in \Sigma$
above $x = \hat{\pi}(P)$, normalised by
\begin{equation} \label{h normalisation}
\bm{\alpha} \cdot \bm{h}(P) = 1,
\end{equation}
where we choose here\footnote{In contrast to \cite{Paper1} where
the normalisation $\bm{\alpha} = (1,0)$ was used. Changing the
normalisation of $\bm{h}$ will obviously have no effect on the
reconstructed solution since this is constructed out of a vector
proportional to $\bm{h}$ anyway.} $\bm{\alpha} = (1,1)$ following
\cite{Krichever:2001zg}. Its components are meromorphic functions
on $\Sigma$ and it follows from a standard argument that
$\bm{h}(P,\sigma,\tau)$ has $g + 1$ poles\footnote{A vector
$\bm{v}(P)$ on $\Sigma$ is said to have a pole at $Q \in \Sigma$
if at least one of its components has a pole at $Q$.} on $\Sigma$
in the present case, which we denote by
$\hat{\gamma}(\sigma,\tau)$.
At this point we can fix some of the gauge redundancy of
\eqref{zero-curvature} by using the gauge transformation $\bm{h}
\mapsto H(\infty)^{-1} \bm{h}$ where $H(x) = \left( \bm{h}(x^+),
\bm{h}(x^-)\right)$ to set $\bm{h}(\infty^+) = {\tiny
\left(\begin{array}{c} 1 \\ 0 \end{array}\right)},
\bm{h}(\infty^-) = {\tiny \left(\begin{array}{c} 0 \\ 1
\end{array}\right)}$; note that this gauge transformation preserves
the normalisation of $\bm{h}$ because $\bm{\alpha}H(x) =
\bm{\alpha}$. The residual gauge symmetry consists of diagonal
matrices $\tilde{g}(\sigma,\tau) = \text{diag}(d_1,d_2)$
\begin{equation} \label{residual gauge symmetry}
\Omega(x) \mapsto \tilde{g} \Omega(x) \tilde{g}^{-1}, \quad \bm{h}
\mapsto f(P)^{-1} \tilde{g} \bm{h}, \quad \text{where } f(P) =
\bm{\alpha} \cdot (\tilde{g} \bm{h}(P)).
\end{equation}
The role of the function $f(P)$ is to keep $\bm{h}$ normalised. It
has the effect of changing the divisor $\hat{\gamma}(\sigma,\tau)$
of poles of $\bm{h}$ to the equivalent divisor
$\hat{\gamma}'(\sigma,\tau)$ ($\sim \hat{\gamma}(\sigma,\tau)$) of
zeroes of $f$. Given a divisor $\hat{\gamma}(\sigma,\tau)$ of
degree $g+1$, the following analytic properties uniquely specify
the components of $\bm{h}$ by the Riemann-Roch theorem
\begin{equation*}
(h_1) \geq \hat{\gamma}(\sigma,\tau)^{-1} \infty^-, \quad
h_1(\infty^+) = 1, \quad \text{ and } \quad (h_2) \geq
\hat{\gamma}(\sigma,\tau)^{-1} \infty^+, \quad h_2(\infty^-) = 1.
\end{equation*}
Note that the divisor of zeroes of $h_1$ is $\gamma(\sigma,\tau)
\infty^-$ where $\gamma(\sigma,\tau)$ denotes the `dynamical
divisor' of degree $g$ that was introduced in \cite{Paper1} (where
$\bm{h}$ was normalised by $\frac{1}{h_1}$ forcing its second
component to have poles at $\gamma(\sigma,\tau) \infty^-$).

The gauge fixing condition $\bm{h}(\infty^+) = {\tiny
\left(\begin{array}{c} 1 \\ 0 \end{array}\right)},
\bm{h}(\infty^-) = {\tiny \left(\begin{array}{c} 0 \\ 1
\end{array}\right)}$ imposed so far also fixes part of the global $SU(2)_R$
symmetry of the equations of motion by restricting the $SU(2)_R$
current $j$ to the level set $Q_R = \frac{1}{2 i} R \sigma_3$.
Indeed, the constant part of \eqref{residual gauge symmetry}
corresponds to the unfixed $U(1)_R$ subgroup of the global
$SU(2)_R$ symmetry group (in fact, before having imposed reality
conditions we are really dealing with a $\mathbb{C}^{\ast}$
subgroup of $SL(2,\mathbb{C})_R$). Let us end this section by
showing that the choice of an initial value for the $U(1)_R$ angle
corresponds exactly to the choice of a representative of the
equivalence class $[\hat{\gamma}(0,0)]$.

Since a specific representative $\hat{\gamma}'(0,0) =
\prod_{i=1}^{g+1} \hat{\gamma}'_i \sim \hat{\gamma}(0,0)$ of the
equivalence class $[\hat{\gamma}(0,0)]$ is uniquely specified by a
single one of its points it suffices to show that for any
arbitrary point $\hat{\gamma}'_1 \in \Sigma$ there exists a unique
diagonal $\tilde{g} \in SL(2,\mathbb{C})_R$ such that
$f(\hat{\gamma}'_1,0,0) = 0$. But a generic diagonal matrix
$\tilde{g} \in SL(2,\mathbb{C})_R$ takes the form
\begin{equation} \label{residual symmetry}
\tilde{g} = \left( \begin{array}{cc} W & 0 \\
0 & W^{-1} \end{array} \right),
\end{equation}
and so the requirement that $f(\hat{\gamma}'_1,0,0) = 0$ simply
reads $W \, h_1(\hat{\gamma}'_1,0,0) + W^{-1} \,
h_2(\hat{\gamma}'_1,0,0) = 0$. The solution to this equation $W^2
= - h_2(\hat{\gamma}'_1,0,0)/h_1(\hat{\gamma}'_1,0,0)$ is unique
(up to a trivial sign) and it follows that $\tilde{g}$ can be
constructed uniquely in such a way that $f(\hat{\gamma}'_1,0,0) =
0$.

For later use we also identify reality conditions on the
representative of the equivalence class $[\hat{\gamma}(0,0)]$, or
more precisely on changes between representatives of
$[\hat{\gamma}(0,0)]$. Given two equivalent divisors
$\hat{\gamma}(0,0)$ and $\hat{\gamma}'(0,0) \sim
\hat{\gamma}(0,0)$ which are the poles and zeroes of $f(P,0,0)$
respectively, the reality requirement $\tilde{g} \in SU(2)_R$
imposes a restriction on the function $f(P,0,0)$ and hence on the
allowed change of divisor $\hat{\gamma}(0,0) \rightarrow
\hat{\gamma}'(0,0)$, namely $|W|^2 = 1$.

\subsubsection{Vector Baker-Akhiezer functions} \label{section: BA vector}

We now look for the analytic properties which uniquely specify the
vector $\bm{\psi}$ solution to the consistency condition $\left( d
- J(x) \right) \bm{\psi} = 0$ of \eqref{zero-curvature}; once the
solution to this system is known, the Lax connection can be
recovered (when $x$ does not correspond to a branch point) by
$J(x) = d\hat{\Psi}(x) \cdot \hat{\Psi}(x)^{-1}$ where
$\hat{\Psi}(x) = (\bm{\psi}(x^+),\bm{\psi}(x^-))$. Since the
operators $d - J(x)$ and $\Omega(x)$ commute we can write
$\bm{\psi}$ as
\begin{equation} \label{psi rep}
\bm{\psi}(P,\sigma,\tau) = \widehat{\Psi}(x,\sigma,\tau)
\bm{h}(P,0,0),
\end{equation}
where $\widehat{\Psi}(x,\sigma,\tau)$ is a formal matrix solution
to $\left( d - J(x) \right) \widehat{\Psi}(x) = 0$. For
definiteness, fix the initial condition to be $\bm{\psi}(P,0,0) =
\bm{h}(P,0,0)$ so that $\widehat{\Psi}(x,0,0) = {\bf 1}$ and hence
$\widehat{\Psi}(x)$ satisfies $\widehat{\Psi}(x,\sigma,\tau)
\Omega(x,0,0) = \Omega(x,\sigma,\tau)
\widehat{\Psi}(x,\sigma,\tau)$. Because $J(x)$ has poles only at
$x = \pm 1$, Poincar\'e's theorem on holomorphic differential
equations implies that $\widehat{\Psi}(x,\sigma,\tau)$ is
holomorphic outside $x = \pm 1$. Its singularities at $x = \pm 1$
are essential singularities of the form \cite{Paper1}
\begin{equation*}
\widehat{\Psi}(x,\sigma,\tau)
e^{-\widehat{S}_{\pm}(x,\sigma,\tau)} = O(1) \quad \text{in a
neighbourhood of } x = \pm 1,
\end{equation*}
where the singular parts $\widehat{S}_{\pm}(x,\sigma,\tau) =
\frac{i \kappa_{\pm}}{2} \frac{\tau \pm \sigma}{1 \mp x} \sigma_3
=: s_{\pm}(x,\sigma,\tau) \sigma_3$ were determined using the
Virasoro constraints. Moreover, using the condition $J(\infty) =
0$ we see that $d \widehat{\Psi}(\infty,\sigma,\tau) = 0$ which
implies $\widehat{\Psi}(\infty,\sigma,\tau) = {\bf 1}$. This is
enough to read off the analytic properties of $\bm{\psi}$ from its
representation in the form \eqref{psi rep} which uniquely specify
its components as Baker-Akhiezer functions, namely
\begin{gather*}
(\psi_1) \geq \hat{\gamma}(0,0)^{-1} \infty^-, \quad \psi_1(\infty^+) =
1, \quad \text{ and } \quad (\psi_2) \geq \hat{\gamma}(0,0)^{-1}
\infty^+, \quad \psi_2(\infty^-) = 1,\\
\text{with } \quad \left\{
\begin{split}
&\psi_i(x^{\pm},\sigma,\tau) e^{\mp s_+(x,\sigma,\tau)} = O(1),
\quad \text{as } x \rightarrow 1,\\
&\psi_i(x^{\pm},\sigma,\tau) e^{\mp s_-(x,\sigma,\tau)} = O(1),
\quad \text{as } x \rightarrow -1.
\end{split}
\right. \notag
\end{gather*}
Given this data which uniquely specifies the Baker-Akhiezer vector
$\bm{\psi}$, one can reconstruct the Lax connection $J(x) =
d\hat{\Psi}(x) \cdot \hat{\Psi}(x)^{-1}$ uniquely up to a residual
gauge transformation. Changing $\hat{\gamma}(0,0)$ to an
equivalent divisor $\hat{\gamma}'(0,0)$ amounts simply to a
scaling $\bm{\psi} \rightarrow k \bm{\psi}$ by a function $k(P)$
with divisor $(k) = \hat{\gamma}(0,0) \cdot
\hat{\gamma}'(0,0)^{-1}$, which has no effect on the reconstructed
Lax connection $J(x) = d\hat{\Psi}(x) \cdot \hat{\Psi}(x)^{-1}$;
therefore the equivalence class $[J(x)]$ of $J(x)$ under residual
gauge transformations is uniquely specified by the equivalence
class $[\hat{\gamma}(0,0)]$ and we have an injective map
\begin{equation} \label{injective map 1}
[J(x)] \mapsto \left\{ \Sigma, dp, [\hat{\gamma}(0,0)] \right\}.
\end{equation}
Since the gauge fixing condition $J(\infty) = 0$ imposed above
still allows for residual gauge transformations by constant
diagonal matrices, corresponding precisely to the unfixed $U(1)_R$
subgroup of the physical symmetry $SU(2)_R$, the initial data
pertaining to the $U(1)_R$ symmetry cannot be determined by
analytical considerations of the auxiliary linear system $\left( d
- J(x) \right)\bm{\psi} = 0$. The best we can do is simplify the
injective map \eqref{injective map 1} down to the following
injective map
\begin{equation} \label{injective map 2}
[j] \mapsto \left\{ \Sigma, dp, [\hat{\gamma}(0,0)] \right\},
\end{equation}
where $[j]$ denotes the equivalence class of $j$ under $U(1)_R$
conjugation. However, the $U(1)_R$ initial angle was argued in
section \ref{section: The normalised eigenvector} to be fully
specified by a choice of representative of the equivalence class
$[\hat{\gamma}(0,0)]$. Thus the full set of initial data of a
finite-gap solution can be completely specified by a divisor
$\hat{\gamma}(0,0)$ of degree $\text{deg }\hat{\gamma}(0,0) =
g+1$. In other words we end up with the following injective map
\begin{equation} \label{injective map 3}
j \mapsto \left\{ \Sigma, dp, \hat{\gamma}(0,0) \right\}.
\end{equation}
This is the analogue of the flat space equation \eqref{flat space
data} in the case at hand of the nonlinear equations of motion for a
string moving on $\mathbb{R} \times S^3$.

This complete set of algebro-geometric data $\left\{ \Sigma, dp,
\hat{\gamma}(0,0) \right\}$ for an arbitrary finite-gap solution $j$
can be succinctly described as a point in the bundle $\mathcal{M}^{(2g
+ 2)}_{\mathbb{C}}$ over $\mathcal{L}$
\begin{equation} \label{bundle M}
S^{g+1}(\Sigma) \rightarrow \mathcal{M}^{(2g + 2)}_{\mathbb{C}}
\rightarrow \mathcal{L},
\end{equation}
whose fibre over every point of the base, specified by a curve
$\Sigma$, is the $(g+1)$-st symmetric product $S^{g+1}(\Sigma) =
\Sigma^{g+1}/S_{g+1}$ of $\Sigma$. If $R$ where to be held fixed
and the global $U(1)_R$ symmetry factored out (as was the case in
\cite{Paper1}), then the leaf would be reduced to $\mathcal{L}|_R$
(see section \ref{section: spectral curve}) and the
$U(1)_R$-reduced solution $[j]$ uniquely specified by the
equivalence class $[\hat{\gamma}(0,0)]$ (see \eqref{injective map
2}) so that the relevant bundle in the $U(1)_R$-reduced case is
\cite{Paper1}
\begin{equation*}
J(\Sigma) \rightarrow \mathcal{M}^{(2g)}_{\mathbb{C}} \rightarrow
\mathcal{L}|_R,
\end{equation*}
using the Abel map $\bm{\mathcal{A}} : S^{g}(\Sigma) \rightarrow
J(\Sigma)$ to identify each fibre with the Jacobian $J(\Sigma)$.

\subsubsection{General finite-gap solution} \label{section: finite-gap integration}

Since the map \eqref{injective map 2} is injective (essentially by the
Riemann-Roch theorem), it admits a left inverse
\begin{equation} \label{injective map 2 inverse}
\left\{ \Sigma, dp, [\hat{\gamma}(0,0)] \right\} \mapsto [j],
\end{equation}
which takes a given set of admissible algebro-geometric data into a
solution of the equations of motion \eqref{eq of motion} and the
Virasoro constraints \eqref{V1}. This solution can be formally
read off from the Lax connection $J(x) = (j - x \ast j)/(1 - x^2)$
constructed out of the Baker-Akhiezer vector $\bm{\psi}$, namely
\begin{equation} \label{Lax reconstruction}
J(x) = d \hat{\Psi}(x) \cdot \hat{\Psi}(x)^{-1} \quad \text{with }
\hat{\Psi}(x) = (\bm{\psi}(x^+),\bm{\psi}(x^-)),
\end{equation}
as already mentioned in the previous section. However, the
algebro-geometric reconstruction of the solution gives more than just
a formal or implicit expression since vector Baker-Akhiezer functions
on a Riemann surface $\Sigma$ admit explicit representations in terms
of Riemann $\theta$-functions associated with $\Sigma$, thus enabling
us to write down explicit formulae for the current $j$.

The analogue of the general flat space solution \eqref{General
solution: flat} with finitely many oscillators turned on, called a
\textit{finite-gap} solution, which solves both \eqref{eq of motion}
and \eqref{V1} was constructed in \cite{Paper1}. Its construction is
based on an algebraic curve $\Sigma$ of finite genus $g$ and is given
by the following expression for the light-cone components of the
current $j$
\begin{multline} \label{General solution: S^3}
j_{\pm}(\sigma,\tau) = e^{\big( \frac{i}{2} \bar{\theta}_0 -
\frac{i}{2} \int_{\infty^-}^{\infty^+} d\mathcal{Q} \big)
\sigma_3} \Theta_{\pm}\left(\bm{\mathcal{A}}(\hat{\gamma}(0,0)) -
\int_{\bm{b}} d\mathcal{Q}; \bm{\mathcal{A}}(\hat{\gamma}(0,0))\right)\\
\times \left( i \kappa_{\pm} \sigma_3 \right)
\Theta_{\pm}\left(\bm{\mathcal{A}}(\hat{\gamma}(0,0)) -
\int_{\bm{b}} d\mathcal{Q};
\bm{\mathcal{A}}(\hat{\gamma}(0,0))\right)^{-1} e^{- \big(
\frac{i}{2}\bar{\theta}_0 - \frac{i}{2} \int_{\infty^-}^{\infty^+}
d\mathcal{Q} \big) \sigma_3},
\end{multline}
where the notation used is defined as follows:
\begin{itemize}
\item[$\bullet$] The differential $d\mathcal{Q}(\sigma,\tau)$ is the
unique normalised second kind Abelian differential with double poles
at the points above $x =\pm 1$ of the prescribed form
\begin{equation*}
d\mathcal{Q} \underset{x \rightarrow \pm 1}\sim i dS_{\pm},
\end{equation*}
where $S_{\pm}(P,\sigma,\tau)$ are the singular parts of the
problem defined as
\begin{equation*}
\left\{
\begin{split}
&S_+(x^{\pm},\sigma,\tau) = \mp \frac{i\kappa_+}{2} \frac{\sigma +
\tau}{x - 1}, \\
&S_-(x^{\pm},\sigma,\tau) = \mp \frac{i\kappa_-}{2} \frac{\sigma -
\tau}{x + 1}.
\end{split}
\right.
\end{equation*}
Note, the matrix $\widehat{S}_{\pm}(x,\sigma,\tau)$ defined in the
pervious section is simply the diagonal matrix $\text{diag}\left(
- S_{\pm}(x^+,\sigma,\tau), - S_{\pm}(x^-,\sigma,\tau) \right)$.

\item[$\bullet$] The divisor $\hat{\gamma}(0,0)$ is the divisor of
poles of $\bm{h}(P,0,0)$ described in the previous sections. Its
degree is $\text{deg } \hat{\gamma}(0,0) = g+1$ and so it lives in
the $(g+1)$-st symmetric product $S^{g+1}(\Sigma) = \Sigma^{g+1} /
S_{g+1}$ of the curve $\Sigma$ which is mapped surjectively onto
the Jacobian $J(\Sigma)$ of $\Sigma$ by means of the Abel map
\begin{equation} \label{Abel map on divisors}
\begin{split}
\bm{\mathcal{A}} : S^{g+1}(\Sigma) &\rightarrow J(\Sigma) \\
\prod_{i=1}^{g+1} P_i &\mapsto 2 \pi \sum_{i=1}^{g+1}
\int_{\infty^+}^{P_i} \bm{\omega}.
\end{split}
\end{equation}

\item[$\bullet$] The solution can only be recovered up to
conjugation by constant diagonal matrices corresponding precisely
to the $\mathbb{C}^{\ast}$ subgroup of $SL(2,\mathbb{C})_R$ (which
becomes the $U(1)_R$ subgroup of $SU(2)_R$ after reality
conditions are imposed) that preserves the level set $Q_R =
\frac{1}{2 i} R \sigma_3$. This undetermined $\mathbb{C}^{\ast}$
conjugation matrix can be expressed in terms of a single arbitrary
constant $\bar{\theta}_0 \in \mathbb{C}$ as $e^{\frac{i}{2}
\bar{\theta}_0 \sigma_3}$. As we have argued, the initial $U(1)_R$
angle $\bar{\theta}_0$ can be specified by the representative
$\hat{\gamma}(0,0)$ of the equivalence class
$\mathcal{A}(\hat{\gamma}(0,0))$. The relation of $\bar{\theta}_0$
to $\hat{\gamma}(0,0)$ will become clear in section \ref{section:
action-angle} when we will extend the target of the Abel
map \eqref{Abel map on divisors} topologically by a $\mathbb{C}^{\ast}$ factor,
turning it into an extended Abel map $\vec{\mathcal{A}} :
S^{g+1}(\Sigma) \rightarrow J(\Sigma, \infty^{\pm})$ that maps
bijectively into the generalised Jacobian $J(\Sigma, \infty^{\pm})$ to
be defined later.

\item[$\bullet$] The function $\Theta_{\pm}$ is $2 \times 2$
matrix valued and its only feature we are interested in for the
present purposes is that its $(\sigma,\tau)$-dependence enters
solely through the $b$-periods of $d\mathcal{Q}(\sigma,\tau)$ in
the expression
\begin{equation*}
\bm{\mathcal{A}}(\hat{\gamma}(\sigma,\tau)) =
\bm{\mathcal{A}}(\hat{\gamma}(0,0)) - \int_{\bm{b}}
d\mathcal{Q}(\sigma,\tau).
\end{equation*}
Likewise, it is important to note that the quantity entering in
the exponents of expression \eqref{General solution: S^3} for
$j_{\pm}$ is just (minus) the $\mathcal{B}_{g+1}$-period of
$d\mathcal{Q}(\sigma,\tau)$, namely
\begin{equation*}
\int_{\infty^-}^{\infty^+} d\mathcal{Q}(\sigma,\tau) =
-\int_{\mathcal{B}_{g+1}} d\mathcal{Q}(\sigma,\tau).
\end{equation*}
\end{itemize}

Moreover, from \eqref{Lax reconstruction} we can also write down a
formal expression for the fundamental field $g$, out of which the
$SU(2)_R$ current $j = - g^{-1} dg = (dg^{-1}) g$ is constructed,
up to an $SU(2)_L$ symmetry (or $SL(2,\mathbb{C})_L$ before
imposing reality conditions)
\begin{equation*}
g = \sqrt{\text{det} \hat{\Psi}(0)} \cdot \hat{\Psi}(0)^{-1} \in
SL(2,\mathbb{C}).
\end{equation*}

As for the current $j$ above, an explicit representations of the
group element $g$ in terms of Riemann $\theta$-functions
associated with $\Sigma$ can also be constructed. Making use of
the dual Baker-Akhiezer vector \cite{Paper1} to express
$\hat{\Psi}(0)^{-1}$ we find that the components $Z_i$ of $g$ in
\eqref{g components} are proportional to the components
$\widetilde{\psi}_i^+(0^+)$ of the dual Baker-Akhiezer vector at
$P = 0^+$, i.e.
\begin{equation*}
\begin{split}
Z_1 = Z_1^0 \frac{\theta \big( 2 \pi \int^{0^+}_{\infty^+}
\bm{\omega} - \int_{\bm{b}} d\mathcal{Q} - \bm{D} \big)}{\theta
\big(\int_{\bm{b}} d\mathcal{Q} + \bm{D} \big)} \; \exp \left( - i
\int_{\infty^+}^{0^+} d\mathcal{Q} \right),\\
Z_2 = Z_2^0 \frac{\theta \big( 2 \pi \int^{0^+}_{\infty^-}
\bm{\omega} - \int_{\bm{b}} d\mathcal{Q} - \bm{D} \big)}{\theta
\big(\int_{\bm{b}} d\mathcal{Q} + \bm{D} \big)} \; \exp \left( - i
\int_{\infty^-}^{0^+} d\mathcal{Q} \right),
\end{split}
\end{equation*}
where $\bm{D} = \bm{\mathcal{A}}(\hat{\gamma}^+(0,0)) +
\bm{\mathcal{K}} \in \mathbb{C}^g$ ($\hat{\gamma}^+(0,0)$ being the
dual divisor to $\hat{\gamma}(0,0)$, see \eqref{dual divisor}, and
$\bm{\mathcal{K}}$ being the vector of Riemann's constants) is almost
arbitrary and $Z_i^0 \in \mathbb{C}$ are constants which can be
expressed in terms of the algebro-geometric data. Using the property
$\hat{\sigma}^{\ast} d\mathcal{Q} = - d\mathcal{Q}$ of the differential
$d\mathcal{Q}$ where $\hat{\sigma} x^{\pm} = x^{\mp}$ is the
hyperelliptic involution of $\Sigma$ we can rewrite the above
expressions in a way that emphasises the linearisation of the motion in
the global $SU(2)_R \times SU(2)_L$ directions, namely\footnote{These
  solutions seem to be closely related to the solutions obtained in
an Appendix of \cite{Kazakov:2004qf} following the method of
\cite{Krichever3}. One apparant difference,
however, is that the latter are constructed from the $\Theta$-functions of
a certain double-cover of the spectral curve $\Sigma$ considered
here. We do not yet understand the precise connection between the two
results.}
\begin{subequations} \label{reconstruction formula for psi^+}
\begin{equation} \label{reconstruction formula for psi^+_1}
Z_1 = Z_1^0 \frac{\theta \big( 2 \pi \int^{0^+}_{\infty^+}
\bm{\omega} - \int_{\bm{b}} d\mathcal{Q} - \bm{D} \big)}{\theta
\big(\int_{\bm{b}} d\mathcal{Q} + \bm{D} \big)} \; \exp \left( + \frac{i}{2}
\int_{\infty^-}^{\infty^+} d\mathcal{Q} -
\frac{i}{2} \int_{0^-}^{0^+} d\mathcal{Q} \right),
\end{equation}
\begin{equation} \label{reconstruction formula for psi^+_2}
Z_2 = Z_2^0 \frac{\theta \big( 2 \pi \int^{0^+}_{\infty^-}
\bm{\omega} - \int_{\bm{b}} d\mathcal{Q} - \bm{D} \big)}{\theta
\big(\int_{\bm{b}} d\mathcal{Q} + \bm{D} \big)} \; \exp \left( - \frac{i}{2}
\int_{\infty^-}^{\infty^+} d\mathcal{Q} -
\frac{i}{2} \int_{0^-}^{0^+} d\mathcal{Q} \right).
\end{equation}
\end{subequations}

\subsection{Extracting data} \label{section: extract data}

Because \eqref{General solution: S^3} is the general solution to
the field equations \eqref{eq of motion}, its restriction to a
given $\tau$-slice, say $\tau = 0$, can be used as a convenient
parametrisation of the most general phase-space configuration
$j(\sigma) = \left( j_0(\sigma), j_1(\sigma) \right)$ of the
string. Furthermore, since the current \eqref{General solution:
S^3} also satisfies the Virasoro constraints \eqref{V1}, the
parameters it depends on are independent physical degrees of freedom
of the string. In the remainder of the paper we shall therefore use
the following parametrisation of the phase-space configuration
$j(\sigma)$
\begin{multline} \label{General configuration: S^3}
j_{\pm}(\sigma) = e^{\big( \frac{i}{2} \bar{\theta}_0 +
\frac{i}{2} n_{g+1} \sigma \big) \sigma_3}
\Theta_{\pm}\left(\bm{\mathcal{A}}(\hat{\gamma}(0,0)) -
\bm{k} \sigma; \bm{\mathcal{A}}(\hat{\gamma}(0,0))\right)\\
\times \left( i \kappa_{\pm} \sigma_3 \right)
\Theta_{\pm}\left(\bm{\mathcal{A}}(\hat{\gamma}(0,0)) - \bm{k}
\sigma; \bm{\mathcal{A}}(\hat{\gamma}(0,0))\right)^{-1} e^{- \big(
\frac{i}{2}\bar{\theta}_0 + \frac{i}{2} n_{g+1} \sigma \big)
\sigma_3},
\end{multline}
where $\bm{k} = \int_{\bm{b}} \frac{dp}{2 \pi}$ and $n_{g+1} =
\int_{\mathcal{B}_{g+1}} \frac{dp}{2 \pi}$ after noting that
$d\mathcal{Q}(\sigma,0) = \frac{\sigma}{2 \pi} dp$. This is the
analogue of the mode expansion \eqref{generic PS configuration}
for the general phase-space configuration in the flat-space case.
Just as one can also extract the parameters of the mode expansion
\eqref{extract data: flat} from a general phase-space configuration in
flat-space, it is possible to extract the divisor $\hat{\gamma}(0,0)$
from a general `finite-gap' phase-space configuration \eqref{General
configuration: S^3} as we now show.

Indeed, the divisor $\hat{\gamma}(0,0)$ of poles of $\bm{h}(P,0,0)$
can be extracted \`a la Sklyanin from $\Omega(x) \equiv
\Omega(x,0,0)$. Introducing the notation $\bm{h}_i = \text{res}_{P =
\hat{\gamma}_i} \bm{h}(P,0,0)$ where $\hat{\gamma}(0,0) = \prod_{i =
1}^{g + 1} \hat{\gamma}_i$, we have
\begin{equation} \label{Sklyanin system}
\left\{
\begin{split}
&\Omega(x_{\hat{\gamma}_i}) \bm{h}_i = \Lambda(\hat{\gamma}_i) \bm{h}_i,\\
&\bm{\alpha} \cdot \bm{h}_i = 0.
\end{split}
\right.
\end{equation}
However, to simplify the forthcoming calculations of Poisson
brackets we perform the following similarity transformation on the
system of equations \eqref{Sklyanin system}
\begin{equation} \label{similarity transformation}
\bm{h}_i \mapsto \left( \begin{array}{cc} 1&1\\ 0&1
\end{array} \right) \bm{h}_i = \tilde{\bm{h}}_i, \quad \Omega(x_{\hat{\gamma}_i})
\mapsto \left( \begin{array}{cc} 1&1\\ 0&1
\end{array} \right) \Omega(x_{\hat{\gamma}_i}) \left( \begin{array}{cc} 1&-1\\ 0&1
\end{array} \right) = \widetilde{\Omega}(x_{\hat{\gamma}_i}),
\end{equation}
so that the system now reads
\begin{equation*}
\left\{
\begin{split}
&\widetilde{\Omega}(x_{\hat{\gamma}_i}) \tilde{\bm{h}}_i = \Lambda(\hat{\gamma}_i)
\tilde{\bm{h}}_i,\\
&\left(\tilde{\bm{h}}_i\right)_1 = 0.
\end{split}
\right.
\end{equation*}
The points $\{ \hat{\gamma}_i \}_{i = 1}^{g + 1}$ of the divisor
$\hat{\gamma}(0,0)$ are therefore characterised in terms of the
components $\widetilde{\mathcal{A}}(x)$ and
$\widetilde{\mathcal{B}}(x)$ of
\begin{equation} \label{tildeOmega components}
\widetilde{\Omega}(x) = \left(
\begin{array}{cc} \widetilde{\mathcal{A}}(x) & \widetilde{\mathcal{B}}(x)\\
\widetilde{\mathcal{C}}(x) & \widetilde{\mathcal{D}}(x)
\end{array} \right) = \left( \begin{array}{cc} 1&1\\ 0&1
\end{array} \right) \left(
\begin{array}{cc} \mathcal{A}(x) & \mathcal{B}(x)\\
\mathcal{C}(x) & \mathcal{D}(x)
\end{array} \right) \left( \begin{array}{cc} 1&-1\\ 0&1
\end{array} \right),
\end{equation}
as follows
\begin{subequations} \label{extracting oscillators: S^3}
\begin{equation} \label{extracting oscillators: S^3 a}
\widetilde{\mathcal{B}}(x_{\hat{\gamma}_i}) = 0, \quad
\Lambda(\hat{\gamma}_i) =
\widetilde{\mathcal{D}}(x_{\hat{\gamma}_i}) =
\widetilde{\mathcal{A}}(x_{\hat{\gamma}_i})^{-1}.
\end{equation}
Note that $\widetilde{\mathcal{B}}(x)$ actually has infinitely
many zeroes but only $g+1$ of them constitute the divisor
$\hat{\gamma}(0,0)$, the remaining zeroes being the singular
points of the curve $\Gamma$. Thus the initial data
$\hat{\gamma}(0,0)$ pertaining to the divisor
$\hat{\gamma}(\sigma,\tau)$, i.e. to the physical degrees of
freedom, can be retrieved from the $\widetilde{\mathcal{A}}$ and
$\widetilde{\mathcal{B}}$ components of
\begin{equation} \label{extracting oscillators: S^3 b}
\widetilde{\Omega}(x) = \left( \begin{array}{cc} 1&1\\ 0&1
\end{array} \right) P \overleftarrow{\exp} \int_0^{2 \pi}
d\sigma' \frac{1}{2} \left( \frac{j_+(\sigma')}{1 - x} -
\frac{j_-(\sigma')}{1 + x} \right) \left( \begin{array}{cc}
1&-1\\ 0&1 \end{array} \right).
\end{equation}
\end{subequations}
Equations \eqref{extracting oscillators: S^3 a} and
\eqref{extracting oscillators: S^3 b} will be our way of
extracting the initial data pertaining to the physical degrees of
freedom from a general field configuration. This is the non-linear
analogue of extracting the Fourier coefficients in the flat space
case, c.f. equation \eqref{extract data: flat}.

Note that the matrix from which one reads off the divisor
$\hat{\gamma}(0,0)$ isn't exactly the monodromy matrix $\Omega(x)$
itself but instead a similarity transformation of it, namely
\begin{equation*}
\widetilde{\Omega}(x) = \left( \begin{array}{cc} 1 &1\\ 0 &1
\end{array} \right) \Omega(x) \left( \begin{array}{cc} 1 &-1\\ 0 &1
\end{array} \right).
\end{equation*}
Therefore in Appendix \ref{section: SL invariance} we relate the
bracket $\{ \widetilde{\Omega}(x) \overset{\otimes},
\widetilde{\Omega}(x') \}$ to the bracket \eqref{fundamental
Poisson bracket} of monodromy matrices. The result is simply that
the matrix $\widetilde{\Omega}(x)$ satisfies exactly the same
algebra as the monodromy matrix $\Omega(x)$ itself, and so we
shall henceforth only refer to $\widetilde{\Omega}(x)$ since it is
the matrix relevant for retrieving the divisor
$\hat{\gamma}(0,0)$.

\subsection{Poisson brackets of algebro-geometric data} \label{section: PB data 2}

Poisson brackets between the components
$\widetilde{\mathcal{A}}(x)$ and $\widetilde{\mathcal{B}}(x)$ of
$\widetilde{\Omega}(x)$ can be deduced from \eqref{fundamental
Poisson bracket} as is done in Appendix \ref{section: components
PB},
\begin{subequations} \label{A and B PB}
\begin{equation} \label{PB1}
\left\{ \widetilde{\mathcal{A}}(x),\widetilde{\mathcal{A}}(x')
\right\} = \left( \widetilde{\mathcal{B}}(x)
\widetilde{\mathcal{C}}(x') - \widetilde{\mathcal{B}}(x')
\widetilde{\mathcal{C}}(x) \right) \hat{s}(x,x'),
\end{equation}
\begin{equation} \label{PB2}
\begin{split}
\left\{ \widetilde{\mathcal{A}}(x),\widetilde{\mathcal{B}}(x')
\right\} = \left(
\widetilde{\mathcal{A}}(x)\widetilde{\mathcal{B}}(x') +
\widetilde{\mathcal{A}}(x')\widetilde{\mathcal{B}}(x) \right) \hat{r}(x,x') \\
+ \left( \widetilde{\mathcal{A}}(x)\widetilde{\mathcal{B}}(x') +
\widetilde{\mathcal{D}}(x')\widetilde{\mathcal{B}}(x) \right)
\hat{s}(x,x'),
\end{split}
\end{equation}
\begin{equation} \label{PB3}
\left\{ \widetilde{\mathcal{B}}(x),\widetilde{\mathcal{B}}(x')
\right\} = 0,
\end{equation}
\end{subequations}
where $\hat{r}(x,x')$ and $\hat{s}(x,x')$ are defined as $r(x,x')$
and $s(x,x')$ respectively without the factors of $\eta$, i.e.
$r(x,x') = \hat{r}(x,x') \eta$ and $s(x,x') = \hat{s}(x,x') \eta$.

In this subsection, we will show that the above relations imply
non-trivial Poisson brackets between the complex variables
comprising the algebro-geometric data.  We will consider the
implications of the three relations \eqref{A and B PB} in turn.
First we take the limit $x' \rightarrow x_{\hat{\gamma}_l}$ of
\eqref{PB1}. Using \eqref{extracting oscillators: S^3} this gives
\begin{equation}\label{PB4}
\{ \widetilde{\mathcal{A}}(x), \Lambda(\hat{\gamma}_l)^{-1} \} =
\widetilde{\mathcal{B}}(x)
\widetilde{\mathcal{C}}(x_{\hat{\gamma}_l})\hat{s}\left(x,x_{\hat{\gamma}_l}\right).
\end{equation}
Taking the limit $x\rightarrow x_{\hat{\gamma}_k}$ immediately
gives,
\begin{equation*}
\{ \Lambda(\hat{\gamma}_k)^{-1}, \Lambda(\hat{\gamma}_l)^{-1} \} =
0.
\end{equation*}

We now turn to the Poisson bracket \eqref{PB2}. Taking the limit
$x \rightarrow x_{\hat{\gamma}_l}$ first gets rid of the terms
proportional to $\widetilde{\mathcal{B}}(x)$ (using
$\widetilde{\mathcal{B}}(x_{\hat{\gamma}_l}) = 0$) and leaves
\begin{equation*}
\left\{
\widetilde{\mathcal{A}}(x_{\hat{\gamma}_l}),\widetilde{\mathcal{B}}(x')
\right\} =
\widetilde{\mathcal{A}}(x_{\hat{\gamma}_l})\widetilde{\mathcal{B}}(x')
\left( \hat{r}(x_{\hat{\gamma}_l},x') +
\hat{s}(x_{\hat{\gamma}_l},x') \right).
\end{equation*}
Now using \eqref{extracting oscillators: S^3} we can write
$\widetilde{\mathcal{B}}(x') = (x' - x_{\hat{\gamma}_k})
\widetilde{\mathcal{B}}_k(x')$ with
$\widetilde{\mathcal{B}}_k(x_{\hat{\gamma}_k}) \neq 0$, so that
\begin{multline*}
(x' - x_{\hat{\gamma}_k}) \left\{
\widetilde{\mathcal{A}}(x_{\hat{\gamma}_l}),\widetilde{\mathcal{B}}_k(x')
\right\} - \left\{
\widetilde{\mathcal{A}}(x_{\hat{\gamma}_l}),x_{\hat{\gamma}_k}
\right\} \widetilde{\mathcal{B}}_k(x') \\ =
\widetilde{\mathcal{A}}(x_{\hat{\gamma}_l}) (x' -
x_{\hat{\gamma}_k}) \widetilde{\mathcal{B}}_k(x') \left(
\hat{r}(x_{\hat{\gamma}_l},x') +
\hat{s}(x_{\hat{\gamma}_l},x')\right),
\end{multline*}
where
\begin{equation*}
\hat{r}(x_{\hat{\gamma}_l},x') + \hat{s}(x_{\hat{\gamma}_l},x') =
- \frac{2 \pi}{\sqrt{\lambda}} \frac{x_{\hat{\gamma}_l}^2 + x'^2 -
2 x_{\hat{\gamma}_l}^2 x'^2}{(x_{\hat{\gamma}_l} - x')(1 -
x_{\hat{\gamma}_l}^2)(1 - x'^2)} - \frac{2 \pi}{\sqrt{\lambda}}
\frac{x_{\hat{\gamma}_l} + x'}{(1 - x_{\hat{\gamma}_l}^2)(1 -
x'^2)}.
\end{equation*}
It is easy to see that taking the limit $x' \rightarrow
x_{\hat{\gamma}_k}$ with $k \neq l$ kills everything but the
second term on the left hand side, leaving $\{
\widetilde{\mathcal{A}}(x_{\hat{\gamma}_l}), x_{\hat{\gamma}_k} \}
= 0, \, k \neq l$. Now setting $k=l$ and taking the limit $x'
\rightarrow x_{\hat{\gamma}_l}$ kills the $\hat{s}$ term leaving
$-\left\{ \widetilde{\mathcal{A}}(x_{\hat{\gamma}_l}),
x_{\hat{\gamma}_l} \right\} = \frac{4 \pi}{\sqrt{\lambda}}
\widetilde{\mathcal{A}}(x_{\hat{\gamma}_l})
\frac{x_{\hat{\gamma}_l}^2}{1 - x_{\hat{\gamma}_l}^2}$. Thus,
again using \eqref{extracting oscillators: S^3}, we have
\begin{equation*}
\left\{ \Lambda(\hat{\gamma}_l)^{-1}, x_{\hat{\gamma}_k} \right\}
= \frac{4 \pi}{\sqrt{\lambda}} \Lambda(\hat{\gamma}_l)^{-1}
\frac{x_{\hat{\gamma}_l}^2}{x_{\hat{\gamma}_l}^2 - 1} \delta_{kl}.
\end{equation*}

Finally we turn our attention to \eqref{PB3}. Again, writing
$\widetilde{\mathcal{B}}(x) = (x - x_{\hat{\gamma}_l})
\widetilde{\mathcal{B}}_l(x)$ it immediately follows from the
third equation \eqref{PB3} that $\left\{ x_{\hat{\gamma}_l},
\widetilde{\mathcal{B}}(x') \right\} = 0$ which in turn implies
that for all $k, l = 1 \ldots, g+1$
\begin{equation*}
\left\{ x_{\hat{\gamma}_l}, x_{\hat{\gamma}_k} \right\} = 0.
\end{equation*}

The algebro-geometric data needed to reconstruct a finite-gap
solution is specified by the $2K=2(g+1)$ complex coordinates,
$\left\{ x_{\hat{\gamma}_l}, \Lambda(\hat{\gamma}_l) \right\}$ for
$l=1, \ldots, g+1$. The results obtained above constitute a
complete set of Poisson brackets for these variables. To write
these brackets in canonical form we change variables to,
\begin{equation*}
z(\hat{\gamma}_l) = x_{\hat{\gamma}_l} +
\frac{1}{x_{\hat{\gamma}_l}}, \qquad \mathcal{P}(\hat{\gamma}_l) =
\frac{\sqrt{\lambda}}{4 \pi} \log\Lambda(\hat{\gamma}_l).
\end{equation*}
Note that $\mathcal{P}(\gamma_l)$ is related to the quasi-momentum
at the point $\hat{\gamma}_l$ as $\mathcal{P}(\hat{\gamma}_l) =
\frac{i \sqrt{\lambda}}{4 \pi} p(\hat{\gamma}_l)$.

In these variables the complete set of Poisson brackets for the
algebro-geometric data becomes,
\begin{subequations} \label{symplectic structure}
\begin{equation}
\label{symplectic structure (a)} \left\{ z(\hat{\gamma}_l),
z(\hat{\gamma}_m) \right\} = 0,
\end{equation}
\begin{equation} \label{symplectic structure (b)}
\left\{ \mathcal{P}(\hat{\gamma}_l), \mathcal{P}(\hat{\gamma}_m)
\right\} = 0,
\end{equation}
\begin{equation} \label{symplectic structure (c)}
\left\{ z(\hat{\gamma}_l), \mathcal{P}(\hat{\gamma}_m) \right\} =
\delta_{lm}.
\end{equation}
\end{subequations}

\subsection{Action-angle variables} \label{section: action-angle}

The change of coordinates to action-angle variables is fairly
standard and was reviewed in the case of the internal degrees of
freedom of the string in \cite{Paper1}. Here we construct the
complete set of action-angle variables starting from the
algebro-geometric symplectic form \eqref{symplectic structure} on
$\mathcal{M}_{\mathbb{C}}^{(2g+2)}$ obtained in the previous
section,
\begin{equation} \label{alg-geom sympl form}
\hat{\omega}_{2K} = - \frac{\sqrt{\lambda}}{4 \pi i}
\sum_{i=1}^{g+1} \delta p(\hat{\gamma}_i) \wedge \delta
z(\hat{\gamma}_i),
\end{equation}
which is naturally defined on the symmetric product bundle
$\mathcal{M}_{\mathbb{C}}^{(2g+2)}$ over $\mathcal{L}$ introduced
in \eqref{bundle M}.

\subsubsection{Symplectic transformation}

It is useful to consider first the universal curve bundle
$\mathcal{N}$ over the leaf $\mathcal{L}$
\begin{equation*}
\Sigma \rightarrow \mathcal{N} \rightarrow \mathcal{L},
\end{equation*}
whose fibre over every point of the base $\mathcal{L}$ is the
corresponding curve $\Sigma$. Recall from section \ref{section:
spectral curve} that the $\{ S_i \}_{i=1}^g$ and $R$ defined in
\eqref{independent moduli} form a set of coordinates on the base
$\mathcal{L}$, and note that $z$ can be taken as a coordinate
along the fibre. Denote by $\delta$ the exterior derivative on the
total space $\mathcal{N}$ and consider the differential $\delta
\tilde{\alpha}$ of $\tilde{\alpha} = - \frac{\sqrt{\lambda}}{4 \pi
i} p dz$ on $\mathcal{N}$
\begin{equation} \label{dpdz}
- \frac{\sqrt{\lambda}}{4 \pi i} \delta p \wedge dz = \delta
\tilde{\alpha} = \sum_{i=1}^g \delta S_i \wedge \partial_{S_i}
\tilde{\alpha} + \frac{1}{2} \delta R \wedge
\partial_{\frac{R}{2}} \tilde{\alpha}.
\end{equation}
The coordinates $\{ S_i \}_{i=1}^g$ and $R$ can be expressed in
terms as the appropriately normalised $\bm{a}$-periods and residue
at $\infty^+$ of the differential $\tilde{\alpha}$,
\begin{equation} \label{action variables}
S_i = \frac{1}{2 \pi} \int_{a_i} \tilde{\alpha}, \; i = 1, \ldots,
g, \quad \frac{R}{2} = - \frac{1}{2 \pi} \int_{c_{\infty^+}}
\tilde{\alpha},
\end{equation}
where $c_{\infty^+}$ is a counter-clockwise cycle around the point
$\infty^+ \in \Sigma$. Now the key observation is that although
$\tilde{\alpha}$ is neither single-valued nor holomorphic on $\Sigma$,
the ambiguities in its definition are constant along the leaf
$\mathcal{L}$ and its pole parts are constant except for those around
$\infty^{\pm} \in \Sigma$ which are proportional to $R$.
It follows therefore from \eqref{action variables} that
\begin{equation*}
\partial_{S_i} \tilde{\alpha} = 2 \pi \omega_i, \; i = 1, \ldots, g, \quad
\partial_{\frac{R}{2}} \tilde{\alpha} = - 2 \pi \omega_{\infty},
\end{equation*}
where $\omega_{\infty}$ is the normalised Abelian differential of
the third kind with simple poles at $\infty^{\pm}$ with residues
$\pm \frac{1}{2 \pi i}$ respectively. Therefore \eqref{dpdz}
simplifies to
\begin{equation*}
\delta \tilde{\alpha} = \sum_{i=1}^g \delta S_i \wedge 2 \pi
\omega_i - \frac{1}{2} \delta R \wedge 2 \pi \omega_{\infty}.
\end{equation*}

This differential $\delta \tilde{\alpha}$ living on $\mathcal{N}$
can be used to define the symplectic form $\hat{\omega}_{2K}$ on
$\mathcal{M}^{(2g+2)}_{\mathbb{C}}$ by the following expression
which is symmetric in the points $\hat{\gamma}_j \in \Sigma$, $j =
1, \ldots, g+1$,
\begin{equation*}
\hat{\omega}_{2K} = \sum_{j = 1}^{g+1} \delta
\tilde{\alpha}(\hat{\gamma}_j) = \sum_{i=1}^g \delta S_i \wedge 2
\pi \left( \sum_{j = 1}^{g+1} \omega_i(\hat{\gamma}_j) \right) -
\frac{1}{2} \delta R \wedge 2 \pi \left( \sum_{j = 1}^{g+1}
\omega_{\infty}(\hat{\gamma}_j) \right).
\end{equation*}
However, the $(g+1)^{\text{st}}$ symmetric product
$S^{g+1}(\Sigma) = \Sigma^{g+1}/S_{g+1}$ of the curve $\Sigma$ is
isomorphic to the $(g+1)$-dimensional \textit{generalised
Jacobian}\footnote{The generalised Jacobian is an extension of the
standard notion of a Jacobian to singular surfaces (see for example
\cite{Fedorov} and references therein) which can be thought of as
limits of regular Riemann surfaces. In the present case the
singular curve is $\Sigma / \{ \infty^{\pm}\}$ (with a degenerated
handle at $\infty$) and its generalised Jacobian $J(\Sigma,
\infty^{\pm})$ is topologically equivalent to the Cartesian product
$J(\Sigma) \times \mathbb{C}^{\ast}$ of the standard $g$-dimensional
Jacobian $J(\Sigma)$ with the cylinder $\mathbb{C}^{\ast} = \mathbb{C}
\setminus \{ 0 \}$.} $J(\Sigma, \infty^{\pm})$ of the curve $\Sigma$
with two punctures at $\infty^{\pm}$ via the extended Abel map
\begin{equation} \label{extended Abel map}
\begin{split}
\vec{\mathcal{A}} : S^{g+1}(\Sigma) &\rightarrow J(\Sigma, \infty^{\pm})\\
D = \prod_{j=1}^{g+1} P_j &\mapsto \left( \bm{\mathcal{A}}(D),
\mathcal{A}_{g+1}(D) \right) = \left( 2 \pi \sum_{j=1}^{g+1}
\int^{P_j}_{P_0} \bm{\omega}, - 2 \pi \sum_{j=1}^{g+1}
\int^{P_j}_{P_0} \omega_{\infty} \right),
\end{split}
\end{equation}
where $P_0 \in \Sigma$ is arbitrary. The first $g$ components of
this map make up the usual Abel map $\bm{\mathcal{A}} :
S^{g+1}(\Sigma) \rightarrow J(\Sigma)$ defined in \eqref{Abel map
on divisors} on divisors of degree $g+1$. Whereas the Abel map
\eqref{Abel map on divisors} was surjective, the extended Abel map
\eqref{extended Abel map} is bijective.

So if we define (complex) coordinates on $J(\Sigma, \infty^{\pm})$ as
\begin{equation} \label{angle variables}
\bm{\theta} = \bm{\mathcal{A}}(\hat{\gamma}(0,0)), \quad
\bar{\theta} = \mathcal{A}_{g+1}(\hat{\gamma}(0,0))
\end{equation}
and identify $\mathcal{M}_{\mathbb{C}}^{(2g+2)}$ with the extended
Jacobian bundle $J(\Sigma, \infty^{\pm}) \rightarrow
\mathcal{J}(\Sigma) \rightarrow \mathcal{L}$ using the extended
Abel map \eqref{extended Abel map} then
\begin{equation} \label{action-angle bracket}
\hat{\omega}_{2K} = \sum_{i=1}^g \delta S_i \wedge \delta \theta_i
+ \frac{1}{2} \delta R \wedge \delta \bar{\theta}.
\end{equation}
It will be convenient to consider a slightly different set of
action-angle variables first proposed in \cite{Paper1} in which the
filling fractions \eqref{filling fractions} play the role of the
action variables. For this we rewrite \eqref{action-angle bracket} as
follows
\begin{equation} \label{action-angle bracket 2}
\begin{split}
\hat{\omega}_{2K} &= \sum_{i=1}^g \delta S_i \wedge \left(
\delta \theta_i - \delta \bar{\theta} \right) + \left( \frac{1}{2} \delta R +
\sum_{i=1}^g \delta S_i \right) \wedge \delta \bar{\theta}\\
&= \sum_{i=1}^g \delta S_i \wedge \delta \left(
\theta_i - \bar{\theta} \right) + \delta \left( \frac{L - R}{2} -
\sum_{i=1}^g S_i \right) \wedge \delta \left( - \bar{\theta} \right),
\end{split}
\end{equation}
where in the second line we use the fact that $\delta L = 0$ since $L$
is fixed along the leaf $\mathcal{L}$ under consideration. Now
recalling the definition of the $K = g+1$ filling fractions $\{
\mathcal{S}_I \}_{I = 1}^K$ introduced in \eqref{filling fractions}
and introducing a new set of angle variables $\{ \varphi_I \}_{I =
1}^K$ related to the $\bm{\theta}, \bar{\theta}$ by \cite{Paper1}
\begin{equation*}
\varphi_I = \theta_i - \bar{\theta} \quad \text{for} \; I = i = 1,
\ldots, g = K-1, \qquad \varphi_K = - \bar{\theta},
\end{equation*}
then equation \eqref{action-angle bracket 2} reads
\begin{equation} \label{action-angle bracket 3}
\hat{\omega}_{2K} = \sum_{I=1}^K \delta \mathcal{S}_I \wedge \delta \varphi_I.
\end{equation}
This of course implies the desired canonical Poisson brackets between
action-angle variables, the non-trivial ones being
\begin{equation*}
\left\{ \mathcal{S}_I, \varphi_J \right\} = \delta_{IJ}, \quad I,J =
1, \ldots, K.
\end{equation*}
Now since we are working in the reduced phase-space after having
imposed the Virasoro constraints and the static gauge fixing
conditions, the relevant bracket to consider for the reduced system is
the Dirac bracket. However we argue in appendix \ref{section: Dirac
brackets} that the Dirac brackets of the action-angle variables are
in fact identical to their Poisson bracket counterparts, so the following
canonical structure also holds in terms of Dirac brackets,
\begin{equation*}
\left\{ \mathcal{S}_I, \varphi_J \right\}_D = \delta_{IJ}, \quad I,J =
1, \ldots, K.
\end{equation*}

\subsubsection{Reality conditions} \label{section: reality}

The reality of the action variables \eqref{action variables}
\begin{equation*}
\bar{S}_i = S_i, \; i = 1, \ldots, g, \quad \bar{R} = R,
\end{equation*}
follows immediately \cite{Paper1} from the reality conditions
$\overline{\hat{\tau}^{\ast} \tilde{\alpha}} = - \tilde{\alpha},
\hat{\tau} \bm{a} = - \bm{a}$ and $\hat{\tau} c_{\infty^+} = -
c_{\infty^+}$ on the $1$-form $\tilde{\alpha}$, the
$\bm{a}$-periods and the cycle $c_{\infty^+}$.

Obtaining real angle variables \eqref{angle variables} is slightly
more involved. In fact, as defined in \eqref{angle variables} the
angles $\bm{\theta}, \bar{\theta}$ are not real but can be made
real after substraction of a constant in each case, which does not
affect the result \eqref{action-angle bracket} of the previous
section. We start by recalling from \cite{Paper1} how to obtain
real angles $\bm{\theta}$.

The equivalence class $[\hat{\gamma}(0,0)]$ of the degree $g+1$
divisor $\hat{\gamma}(0,0)$ satisfies a simple reality condition
\cite{Paper1}, namely
\begin{equation} \label{divisor reality}
\hat{\tau} \hat{\gamma}(0,0) \sim \hat{\gamma}^+(0,0),
\end{equation}
where $\hat{\gamma}^+(0,0)$ is the dual divisor to
$\hat{\gamma}(0,0)$. The dual divisor is of degree $g+1$ and can
be defined up to equivalence by the relation
\begin{equation} \label{dual divisor}
\hat{\gamma}(0,0) \cdot \hat{\gamma}^+(0,0) \sim Z \cdot
(\infty^+)^2 \cdot (\infty^-)^2,
\end{equation}
where $Z$ is the canonical class, i.e. the divisor of any Abelian
differential (the ratio of any two Abelian differential is a
meromorphic function and so their divisors are equivalent).
Putting equations \eqref{divisor reality} and \eqref{dual divisor}
together, the reality condition on $[\hat{\gamma}(0,0)]$ can be
expressed as follows
\begin{equation} \label{divisor reality 2}
\hat{\gamma}(0,0) \cdot \hat{\tau} \hat{\gamma}(0,0) \sim Z \cdot
(\infty^+)^2 \cdot (\infty^-)^2.
\end{equation}
If the base point $P_0$ of the Abel map is chosen to be real, i.e.
such that $\hat{\tau} P_0 = P_0$, then the reality condition on the
Abel map reads $\bm{\mathcal{A}}(\hat{\tau} D) = -
\overline{\bm{\mathcal{A}}(D)}$. It follows then from
\eqref{divisor reality 2} that
\begin{equation*}
2 \; \text{Im } \bm{\mathcal{A}}(\hat{\gamma}(0,0)) =
\bm{\mathcal{A}} \left(Z \cdot (\infty^+)^2 \cdot (\infty^-)^2
\right).
\end{equation*}
This yields the reality condition on the first $g$ components
$\bm{\mathcal{A}}(\hat{\gamma}(0,0))$ of
$\vec{\mathcal{A}}(\hat{\gamma}(0,0))$. The angle coordinates
$\bm{\theta}$ are rendered real after the following redefinition
\begin{equation} \label{real angle variables 1}
\bm{\theta} = \bm{\mathcal{A}}(\hat{\gamma}(0,0)) - \frac{1}{2}
\bm{\mathcal{A}}(Z \cdot (\infty^+)^2 \cdot (\infty^-)^2) \in
\text{Re } J(\Sigma).
\end{equation}

We now turn to the reality of the angle $\bar{\theta}$. For this
we show that under a change of representative of the class
$[\hat{\gamma}(0,0)]$ which is such that the reality condition
$|W|^2 = 1$ on \eqref{residual symmetry} is satisfied, the
corresponding change $\Delta \bar{\theta}$ in the angle $\bar{\theta}$
is real. It would follow from this that representatives of
$[\hat{\gamma}(0,0)]$ which give rise to real solutions are mapped
under $\mathcal{A}_{g+1}$ to a real subspace of $\mathbb{C}$,
i.e. $\mathcal{A}_{g+1}(\hat{\gamma}(0,0)) - \mathcal{C} \in
\mathbb{R}$ for some constant $\mathcal{C} \in \mathbb{C}$.

Recall the function $f(P,0,0) = W h_1(P,0,0) + W^{-1} h_2(P,0,0)$
introduced in \eqref{residual gauge symmetry} which has poles at
$\hat{\gamma}(0,0)$ and zeroes at the equivalent divisor
$\hat{\gamma}'(0,0) = \prod_{i=1}^{g+1} \hat{\gamma}'_i \sim
\hat{\gamma}(0,0)$, and consider the differential $df/f = d \,
\text{log} \,f$. Its only poles are at $\hat{\gamma}(0,0)$ with
residues $-1$ and at $\hat{\gamma}'(0,0)$ with residues $+1$. For
an arbitrary pair of points $P,Q \in \Sigma$, let us denote by
$\omega_{PQ}$ the unique normalised (vanishing $\bm{a}$-periods)
Abelian differential of the third kind with simple poles at $P$
and $Q$ with residues $+1$ and $-1$ there respectively. Then it
follows that
\begin{equation} \label{realthetabar 0}
\frac{df}{f} - \sum_{j=1}^{g+1} \omega_{\hat{\gamma}'_j
\hat{\gamma}_j} = \sum_{i=1}^g c_i \omega_i,
\end{equation}
for some constants $c_i \in \mathbb{C}$. Taking the
$\mathcal{B}_{g+1}$-period of the last equation yields
\begin{equation} \label{realthetabar 1}
\int_{\infty^-}^{\infty^+} \frac{df}{f} - \sum_{j=1}^{g+1}
\int_{\infty^-}^{\infty^+} \omega_{\hat{\gamma}'_j \hat{\gamma}_j}
= \sum_{i=1}^g c_i \int_{\infty^-}^{\infty^+} \omega_i.
\end{equation}
Taking the $\bm{a}$-periods of the equation \eqref{realthetabar 0}
on the other hand gives the constants $c_i$,
\begin{equation*}
c_i = \int_{a_i} \frac{df}{f} = \int_{a_i} d \, \text{log} \, f =
2 \pi i m_i, \quad m_i \in \mathbb{Z}.
\end{equation*}
Now using the Riemann bilinear identities it is straightforward to
show that
\begin{equation} \label{relation between periods}
\int_{\infty^-}^{\infty^+} \omega_{\hat{\gamma}'_j \hat{\gamma}_j}
= 2 \pi i \int_{\hat{\gamma}_j}^{\hat{\gamma}'_j} \omega_{\infty},
\quad \int_{\infty^-}^{\infty^+} \omega_i = \int_{b_i}
\omega_{\infty},
\end{equation}
and so plugging this back into \eqref{realthetabar 1} yields the
following
\begin{equation*}
\int_{\infty^-}^{\infty^+} \frac{df}{f} - 2 \pi i \sum_{j=1}^{g+1}
\int_{\hat{\gamma}_j}^{\hat{\gamma}'_j} \omega_{\infty} = 2 \pi i
\sum_{i=1}^g m_i \int_{b_i} \omega_{\infty}.
\end{equation*}
Referring back to the definition of the extended Abel map
\eqref{extended Abel map} we recognise the second term on the left
hand side of the last expression as the difference between the
$(g+1)^{\text{st}}$ components of the Abel maps of the divisors
$\hat{\gamma}'(0,0)$ and $\hat{\gamma}(0,0)$,
\begin{equation} \label{Delta theta bar}
\Delta \bar{\theta} = \mathcal{A}_{g+1}(\hat{\gamma}'(0,0)) -
\mathcal{A}_{g+1}(\hat{\gamma}(0,0)) = i
\int_{\infty^-}^{\infty^+} \frac{df}{f} + 2 \pi \sum_{i=1}^g m_i
\int_{b_i} \omega_{\infty}.
\end{equation}
So to show that $\Delta \bar{\theta} \in \mathbb{R}$ it suffices to
show the right hand side of \eqref{Delta theta bar} is real. Consider
the first term, which using the limits $f(\infty^{\pm}) = W^{\pm 1}$
can be simplified as
\begin{equation*}
\int_{\infty^-}^{\infty^+} \frac{df}{f} =
\int_{\infty^-}^{\infty^+} d \, \text{log} \, f = \text{log}
\left( \frac{f(\infty^+)}{f(\infty^-)} \right) = 2 \, \text{log}
\, W.
\end{equation*}
This last expression holds as an equality only on $\mathbb{C}/2
\pi i \mathbb{Z}$. We now make use of the reality condition $|W|^2
= 1$ on the residual symmetry \eqref{residual symmetry}, which can
be rewritten $\overline{W} = W^{-1}$, and deduce that
\begin{equation*}
i \int_{\infty^-}^{\infty^+} \frac{df}{f} \in \mathbb{R}/2 \pi
\mathbb{Z}.
\end{equation*}
Furthermore, using \eqref{relation between periods} and the reality
conditions $\overline{\hat{\tau}^{\ast} \bm{\omega}} = - \bm{\omega},
\hat{\tau} \mathcal{B}_{g+1} = \mathcal{B}_{g+1} - \sum_{k=1}^g a_k$
one can show that
\begin{equation*}
\overline{\int_{b_i} \omega_{\infty}} = - \int_{b_i} \omega_{\infty} + 1,
\end{equation*}
so after imposing the reality constraint $|W|^2 = 1$, equation
\eqref{Delta theta bar} implies that
\begin{equation} \label{Delta theta bar 2}
\Delta \bar{\theta} \in \mathbb{R} / 2 \pi \mathbb{Z}.
\end{equation}
As we have already remarked, it follows now from \eqref{Delta theta
bar 2} that the image under $\mathcal{A}_{g+1}$ of the
representatives of $[\hat{\gamma}(0,0)]$ which give rise to real
solutions forms a real subspace of $\mathbb{C}$, i.e.
\begin{equation} \label{real angle variables 2}
\bar{\theta} = \mathcal{A}_{g+1}(\hat{\gamma}(0,0)) - \mathcal{C}
\in \mathbb{R} / 2 \pi \mathbb{Z}.
\end{equation}

\section*{Acknowledgements}
Both authors would like to thank Marc Magro and Jean-Michel Maillet
for interesting discussions, and are very grateful to Marc Magro for
raising the issue of the Dirac brackets. The research of ND is
supported by a PPARC Senior Research Fellowship and BV is supported by
EPSRC.

\appendix

\section{Algebra of transition matrices} \label{section: transition matrices PB}

We start with the Poisson bracket \eqref{Maillet bracket} between
two $J_1(\sigma,x)$ matrices for a non-ultralocal system in the
$(r-s)$-matrix formalism introduced by Maillet \cite{Maillet} which
can be conveniently rewritten as
\begin{align} \label{Maillet bracket 2}
\left\{ J_1(\sigma,x) \mathop{,}^{\otimes} J_1(\sigma',x')\right\}
&= \left[ r(\sigma,x,x'), J_1(\sigma,x)\otimes \mathbf{1} +
\mathbf{1} \otimes J_1(\sigma',x') \right] \delta(\sigma - \sigma')
\notag \\ &- \left[ s(\sigma,x,x'), J_1(\sigma,x)\otimes \mathbf{1}
- \mathbf{1} \otimes J_1(\sigma',x') \right] \delta(\sigma -
\sigma')\\ &- \left( r(\sigma,x,x') + s(\sigma,x,x') -
r(\sigma',x,x') + s(\sigma',x,x') \right)\delta'(\sigma - \sigma'),
\notag
\end{align}
using the identity $\left( f(\sigma) - f(\sigma')
\right)\delta'(\sigma - \sigma') = - f'(\sigma) \delta(\sigma -
\sigma')$ valid for any function $f$. Now the transition matrix
\begin{equation*}
T(\sigma_1,\sigma_2,x) = P \overleftarrow{\exp}
\int_{\sigma_2}^{\sigma_1} d\sigma J_1(\sigma,x),
\end{equation*}
is the unique solution to the following differential equation with
boundary condition
\begin{equation} \label{eq for transition matrix}
\frac{\partial T}{\partial \sigma_1}(\sigma_1,\sigma_2,x) =
J_1(\sigma_1,x) T(\sigma_1,\sigma_2,x), \qquad
T(\sigma_2,\sigma_2,x) = {\bf 1}.
\end{equation}
It also satisfies the following differential equation with the same
boundary condition
\begin{equation*}
\frac{\partial T}{\partial \sigma_2}(\sigma_1,\sigma_2,x) = -
T(\sigma_1,\sigma_2,x) J_1(\sigma_2,x), \qquad
T(\sigma_1,\sigma_1,x) = {\bf 1}.
\end{equation*}
The variation of the system \eqref{eq for transition matrix} gives
\begin{equation*}
\frac{\partial \delta T}{\partial \sigma_1}(\sigma_1,\sigma_2,x) =
\delta J_1(\sigma_1,x) T(\sigma_1,\sigma_2,x) + J_1(\sigma_1,x)
\delta T(\sigma_1,\sigma_2,x), \qquad \delta T(\sigma_1,\sigma_1,x)
= 0,
\end{equation*}
of which the unique solution is easily seen to be
\begin{equation} \label{dT in relation to dJ}
\begin{split}
\delta T(\sigma_1,\sigma_2,x) &= \int_{\sigma_2}^{\sigma_1}
d\sigma T(\sigma_1,\sigma,x) \delta J_1(\sigma,x) T(\sigma,\sigma_2,x), \\
&= \int_{0}^{2 \pi} d\sigma \epsilon(\sigma_1 - \sigma_2)
\chi(\sigma; \sigma_1,\sigma_2) T(\sigma_1,\sigma,x) \delta
J_1(\sigma,x) T(\sigma,\sigma_2,x),
\end{split}
\end{equation}
where $\epsilon(\sigma) = \text{sign}(\sigma)$ is the usual sign
function and $\chi(\sigma; \sigma_1,\sigma_2)$ is the characteristic
function of the interval between $\sigma_1$ and $\sigma_2$.

Now given the Poisson bracket of the system
\begin{equation} \label{PB definition}
\left\{ A \mathop{,}^{\otimes} B \right\} = \int d\sigma \left(
\frac{\delta A}{\delta q^a(\sigma)} \otimes \frac{\delta B}{\delta
\pi^a(\sigma)} - \frac{\delta A}{\delta \pi^a(\sigma)} \otimes
\frac{\delta B}{\delta q^a(\sigma)} \right),
\end{equation}
one can relate the bracket of transition matrices to the bracket of
currents \eqref{Maillet bracket 2} using \eqref{dT in relation to
dJ}
\begin{multline} \label{TT from JJ}
\left\{ T(\sigma_1,\sigma_2,x) \mathop{,}^{\otimes}
T(\sigma'_1,\sigma'_2,x') \right\} = \int_{\sigma_2}^{\sigma_1}
d\sigma \int_{\sigma'_2}^{\sigma'_1} d\sigma' \left(
T(\sigma_1,\sigma,x) \otimes T(\sigma'_1,\sigma',x') \right)\\
\times \left\{ J_1(\sigma,x) \mathop{,}^{\otimes} J_1(\sigma',x')
\right\} \left( T(\sigma,\sigma_2,x) \otimes T(\sigma',\sigma'_2,x')
\right).
\end{multline}
Now plugging \eqref{Maillet bracket 2} into this expression, one
finds after a bit of algebra
\begin{multline*}
\left\{ T(\sigma_1,\sigma_2,x) \mathop{,}^{\otimes}
T(\sigma'_1,\sigma'_2,x') \right\} = \int_0^{2 \pi} d\sigma
\int_0^{2 \pi} d\sigma' \chi(\sigma; \sigma_1,\sigma_2)
\chi(\sigma'; \sigma'_1,\sigma'_2) \epsilon(\sigma_1 - \sigma_2)
\epsilon(\sigma'_1 - \sigma'_2) \\ \times \left[
\frac{\partial}{\partial \sigma} \Big( T(\sigma_1,\sigma,x) \otimes
T(\sigma'_1,\sigma',x') \left( r(\sigma,x,x') - s(\sigma,x,x')
\right) T(\sigma,\sigma_2,x) \otimes
T(\sigma',\sigma'_2,x') \delta(\sigma - \sigma') \Big) \right.\\
+ \left. \frac{\partial}{\partial \sigma'} \Big(
T(\sigma_1,\sigma,x) \otimes T(\sigma'_1,\sigma',x') \left(
r(\sigma,x,x') + s(\sigma,x,x') \right) T(\sigma,\sigma_2,x) \otimes
T(\sigma',\sigma'_2,x') \delta(\sigma - \sigma') \Big) \right].
\end{multline*}
Integrating by parts and using the identity $ -
\frac{\partial}{\partial \sigma} \chi(\sigma; \sigma_1,\sigma_2)
\epsilon(\sigma_1 - \sigma_2) = \delta(\sigma - \sigma_1) -
\delta(\sigma - \sigma_2)$ we obtain
\begin{multline} \label{transition matrix PB}
\left\{ T(\sigma_1,\sigma_2,x) \mathop{,}^{\otimes}
T(\sigma'_1,\sigma'_2,x') \right\} \\
\begin{split}
= &+ \epsilon(\sigma'_1 - \sigma'_2) \chi(\sigma;
\sigma'_1,\sigma'_2)
\\ &\times \left. T(\sigma_1,\sigma,x) \otimes
T(\sigma'_1,\sigma,x') \left( r(\sigma,x,x') - s(\sigma,x,x')
\right) T(\sigma,\sigma_2,x) \otimes T(\sigma,\sigma'_2,x')
\right|_{\sigma = \sigma_2}^{\sigma =
\sigma_1} \\
&+ \epsilon(\sigma_1 - \sigma_2) \chi(\sigma; \sigma_1,\sigma_2)
\\ &\times \left. T(\sigma_1,\sigma,x) \otimes T(\sigma'_1,\sigma,x')
\left( r(\sigma,x,x') + s(\sigma,x,x') \right) T(\sigma,\sigma_2,x)
\otimes T(\sigma,\sigma'_2,x') \right|_{\sigma = \sigma'_2}^{\sigma
= \sigma'_1}
\end{split}
\end{multline}

\section{$SL(2,\mathbb{C})$-invariance of $\left\{ \Omega
\overset{\otimes}, \Omega \right\}$} \label{section: SL
invariance}

In this appendix we wish to find how the Poisson bracket $\left\{
\Omega \overset{\otimes}, \Omega \right\}$ transforms under a
general similarity transformation of $\Omega(x) \rightarrow
\widetilde{\Omega}(x) = U^{-1} \Omega(x) U$, $U \in
SL(2,\mathbb{C})$. Using the shorthand notation $\overset{1}A = A
\otimes {\bf 1}$, $\overset{2}A = {\bf 1} \otimes A$ we can write
\begin{multline*}
\left\{ \overset{1}{\widetilde{\Omega}}(x) ,
\overset{2}{\widetilde{\Omega}}(x') \right\} = \left\{
\overset{1}{U^{-1}} \overset{1}\Omega(x) \overset{1}U ,
\overset{2}{U^{-1}} \overset{2}\Omega(x') \overset{2}U \right\} =
\overset{1}{U^{-1}} \overset{2}{U^{-1}} \left\{
\overset{1}\Omega(x), \overset{2}\Omega(x') \right\} \overset{1}U
\overset{2}U\\ = \overset{1}{U^{-1}} \overset{2}{U^{-1}} \left(
\Big[\overset{12}r(x,x'), \overset{1}\Omega(x)
\overset{2}\Omega(x')\Big] + \overset{1}\Omega(x)
\overset{12}s(x,x') \overset{2}\Omega(x') - \overset{2}\Omega(x')
\overset{12}s(x,x') \overset{1}\Omega(x) \right) \overset{1}U
\overset{2}U\\
= \Big[\overset{12}{\widetilde{r}}(x,x'),
\overset{1}{\widetilde{\Omega}}(x)
\overset{2}{\widetilde{\Omega}}(x')\Big] +
\overset{1}{\widetilde{\Omega}}(x)
\overset{12}{\widetilde{s}}(x,x')
\overset{2}{\widetilde{\Omega}}(x') -
\overset{2}{\widetilde{\Omega}}(x')
\overset{12}{\widetilde{s}}(x,x')
\overset{1}{\widetilde{\Omega}}(x),
\end{multline*}
where $\widetilde{r}(x,x') = U^{-1} \otimes U^{-1} r(x,x') U
\otimes U$ and $\widetilde{s}(x,x') = U^{-1} \otimes U^{-1}
s(x,x') U \otimes U$. Now since $r(x,x')$ and $s(x,x')$ are both
proportional to $\eta = - t^a \otimes t^a$, we can compute the
transformations of $r(x,x')$ and $s(x,x')$ simultaneously by
considering
\begin{equation*}
\left(U^{-1} \otimes U^{-1}\right) \eta \left(U \otimes U\right) =
\left(U^{-1} \otimes {\bf 1}\right) \left({\bf 1} \otimes
U^{-1}\right) \eta \left(U \otimes {\bf 1}\right) \left({\bf 1}
\otimes U\right).
\end{equation*}
Considering an infinitesimal transformation $U = e^{\alpha} \sim
{\bf 1} + \alpha + O(\alpha^2)$, $\alpha \in
\mathfrak{sl}(2,\mathbb{C})$, one finds straightforwardly that
\begin{equation*}
\left(({\bf 1} - \alpha) \otimes ({\bf 1} - \alpha)\right) \eta
\left(({\bf 1} + \alpha) \otimes ({\bf 1} + \alpha)\right) \sim
\eta + O(\alpha^2).
\end{equation*}
Therefore $\eta$ is invariant under infinitesimal similarity
transformations. It follows then that $\widetilde{r}(x,x') =
r(x,x')$ and $\widetilde{s}(x,x') = s(x,x')$, so that the (weak)
bracket $\left\{ \Omega \overset{\otimes}, \Omega \right\}$ ends
up being invariant under similarity transformations as well,
namely the same bracket \eqref{fundamental Poisson bracket} holds
for the transformed monodromy matrix $\widetilde{\Omega}(x)$
\begin{equation} \label{tildeOmega PB}
\begin{split}
\left\{ \widetilde{\Omega}(x) \mathop{,}^{\otimes}
\widetilde{\Omega}(x') \right\} = &[r(x,x'), \widetilde{\Omega}(x)
\otimes \widetilde{\Omega}(x')] \\ +
&\left(\widetilde{\Omega}(x) \otimes {\bf 1}\right) s(x,x') \left(
{\bf 1} \otimes \widetilde{\Omega}(x') \right) \\ - &\left(
{\bf 1} \otimes \widetilde{\Omega}(x') \right) s(x,x') \left(
\widetilde{\Omega}(x) \otimes {\bf 1} \right).
\end{split}
\end{equation}

\section{Algebra of $\widetilde{\mathcal{A}}(x)$ and $\widetilde{\mathcal{B}}(x)$
components} \label{section: components PB}

Let us express the right hand side of \eqref{tildeOmega PB} in
terms of the components \eqref{tildeOmega components} of
$\widetilde{\Omega}(x)$. We have
\begin{equation*}
\begin{array}{c}
\widetilde{\Omega}(x) \otimes \widetilde{\Omega}(x') = \left(
\begin{array}{cc}
\widetilde{\mathcal{A}}(x)\widetilde{\Omega}(x') &
\widetilde{\mathcal{B}}(x)\widetilde{\Omega}(x')\\
\widetilde{\mathcal{C}}(x)\widetilde{\Omega}(x') &
\widetilde{\mathcal{D}}(x)\widetilde{\Omega}(x')
\end{array}
\right), \\ {\bf 1} \otimes \widetilde{\Omega}(x') = \left(
\begin{array}{cc}
\widetilde{\Omega}(x') & 0\\
0 & \widetilde{\Omega}(x')
\end{array}
\right), \widetilde{\Omega}(x) \otimes {\bf 1} = \left(
\begin{array}{cc}
\widetilde{\mathcal{A}}(x) {\bf 1} & \widetilde{\mathcal{B}}(x) {\bf 1}\\
\widetilde{\mathcal{C}}(x) {\bf 1} & \widetilde{\mathcal{D}}(x)
{\bf 1}
\end{array}
\right),
\end{array}
\end{equation*}
and since the matrices $r(x,x')$ and $s(x,x')$ are both proportional
to $\eta$ with
\begin{equation*}
\eta = - t^a \otimes t^a = \frac{1}{2} \sigma_a \otimes \sigma_a =
\frac{1}{2} \left(
\begin{array}{cc}
\sigma_3 & \sigma_1 - i \sigma_2\\
\sigma_1 + i \sigma_2 & -\sigma_3
\end{array}
\right),
\end{equation*}
where $t^a = \frac{i}{\sqrt{2}} \sigma_a$ in the $\mathfrak{su}(2)$
case ($\sigma_a$ being the Pauli matrices), we need to compute the
following quantities
\begin{multline*}
\eta \widetilde{\Omega}(x) \otimes \widetilde{\Omega}(x') \\ =
\frac{1}{2} \left(
\begin{array}{cc}
\widetilde{\mathcal{A}}(x)\sigma_3\widetilde{\Omega}(x') +
\widetilde{\mathcal{C}}(x)(\sigma_1 - i
\sigma_2)\widetilde{\Omega}(x') &
\widetilde{\mathcal{B}}(x)\sigma_3\widetilde{\Omega}(x')
+ \widetilde{\mathcal{D}}(x)(\sigma_1 - i \sigma_2)\widetilde{\Omega}(x')\\
\widetilde{\mathcal{A}}(x)(\sigma_1 + i
\sigma_2)\widetilde{\Omega}(x') -
\widetilde{\mathcal{C}}(x)\sigma_3\widetilde{\Omega}(x') &
\widetilde{\mathcal{B}}(x)(\sigma_1 + i
\sigma_2)\widetilde{\Omega}(x') -
\widetilde{\mathcal{D}}(x)\sigma_3\widetilde{\Omega}(x')
\end{array}\right),
\end{multline*}
\begin{multline*}
\widetilde{\Omega}(x) \otimes \widetilde{\Omega}(x') \eta \\ =
\frac{1}{2} \left(
\begin{array}{cc} \widetilde{\mathcal{A}}(x)\widetilde{\Omega}(x')\sigma_3 +
\widetilde{\mathcal{B}}(x)\widetilde{\Omega}(x')(\sigma_1 + i
\sigma_2) &
\widetilde{\mathcal{A}}(x)\widetilde{\Omega}(x')(\sigma_1 - i
\sigma_2)
- \widetilde{\mathcal{B}}(x)\widetilde{\Omega}(x')\sigma_3\\
\widetilde{\mathcal{C}}(x)\widetilde{\Omega}(x')\sigma_3 +
\widetilde{\mathcal{D}}(x)\widetilde{\Omega}(x')(\sigma_1 + i
\sigma_2) &
\widetilde{\mathcal{C}}(x)\widetilde{\Omega}(x')(\sigma_1 - i
\sigma_2) -
\widetilde{\mathcal{D}}(x)\widetilde{\Omega}(x')\sigma_3
\end{array}
\right),
\end{multline*}
\begin{multline*}
\left( \widetilde{\Omega}(x) \otimes {\bf 1}\right) \eta \left( {\bf 1} \otimes \widetilde{\Omega}(x') \right)\\
= \frac{1}{2} \left(
\begin{array}{cc} \widetilde{\mathcal{A}}(x)\sigma_3 \widetilde{\Omega}(x') +
\widetilde{\mathcal{B}}(x)(\sigma_1 + i
\sigma_2)\widetilde{\Omega}(x') &
\widetilde{\mathcal{A}}(x)(\sigma_1 - i
\sigma_2)\widetilde{\Omega}(x')
- \widetilde{\mathcal{B}}(x)\sigma_3\widetilde{\Omega}(x')\\
\widetilde{\mathcal{C}}(x)\sigma_3 \widetilde{\Omega}(x') +
\widetilde{\mathcal{D}}(x)(\sigma_1 + i
\sigma_2)\widetilde{\Omega}(x') &
\widetilde{\mathcal{C}}(x)(\sigma_1 - i
\sigma_2)\widetilde{\Omega}(x') -
\widetilde{\mathcal{D}}(x)\sigma_3\widetilde{\Omega}(x')
\end{array}
\right),
\end{multline*}
\begin{multline*}
\left( {\bf 1} \otimes \widetilde{\Omega}(x') \right) \eta \left( \widetilde{\Omega}(x) \otimes {\bf 1}\right)\\
= \frac{1}{2} \left(
\begin{array}{cc} \widetilde{\mathcal{A}}(x)\widetilde{\Omega}(x')\sigma_3 +
\widetilde{\mathcal{C}}(x)\widetilde{\Omega}(x')(\sigma_1 - i
\sigma_2) &
\widetilde{\mathcal{B}}(x)\widetilde{\Omega}(x')\sigma_3 +
\widetilde{\mathcal{D}}(x)\widetilde{\Omega}(x')(\sigma_1 - i \sigma_2)\\
\widetilde{\mathcal{A}}(x)\widetilde{\Omega}(x')(\sigma_1 + i
\sigma_2) -
\widetilde{\mathcal{C}}(x)\widetilde{\Omega}(x')\sigma_3 &
\widetilde{\mathcal{B}}(x)\widetilde{\Omega}(x')(\sigma_1 + i
\sigma_2) -
\widetilde{\mathcal{D}}(x)\widetilde{\Omega}(x')\sigma_3
\end{array}
\right).
\end{multline*}
We can read off from this and equation \eqref{tildeOmega PB} the
Poisson brackets between various components of
$\widetilde{\Omega}(x)$, but we are particular interested in the
$\widetilde{\mathcal{A}}(x)$ and $\widetilde{\mathcal{B}}(x)$
components which are given by
\begin{multline*}
\left\{ \widetilde{\mathcal{A}}(x),\widetilde{\mathcal{A}}(x')
\right\} = \left\{
\widetilde{\Omega}_{11}(x),\widetilde{\Omega}_{11}(x') \right\} =
\left\{ \widetilde{\Omega}(x) \mathop{,}^{\otimes}
\widetilde{\Omega}(x') \right\}_{11,11} \\
= \left( \widetilde{\mathcal{B}}(x) \widetilde{\mathcal{C}}(x') -
\widetilde{\mathcal{B}}(x') \widetilde{\mathcal{C}}(x) \right)
\hat{s}(x,x'),
\end{multline*}
where $\hat{s}(x,x') = - \frac{2 \pi}{\sqrt{\lambda}}
\frac{x+x'}{(1 - x^2)(1 - {x'}^2)}$ is $s(x,x')$ without the
matrix factor $\eta$, as well as
\begin{multline*}
\left\{ \widetilde{\mathcal{A}}(x),\widetilde{\mathcal{B}}(x')
\right\} = \left\{
\widetilde{\Omega}_{11}(x),\widetilde{\Omega}_{12}(x') \right\} =
\left\{ \widetilde{\Omega}(x) \mathop{,}^{\otimes}
\widetilde{\Omega}(x') \right\}_{11,12}\\ = \left(
\widetilde{\mathcal{A}}(x)\widetilde{\mathcal{B}}(x') +
\widetilde{\mathcal{A}}(x')\widetilde{\mathcal{B}}(x) \right)
\hat{r}(x,x') + \left(
\widetilde{\mathcal{A}}(x)\widetilde{\mathcal{B}}(x') +
\widetilde{\mathcal{D}}(x')\widetilde{\mathcal{B}}(x) \right)
\hat{s}(x,x'),
\end{multline*}
where $\hat{r}(x,x')$ is $r(x,x')$ without the matrix factor $\eta$,
and lastly
\begin{equation*} \left\{
\widetilde{\mathcal{B}}(x),\widetilde{\mathcal{B}}(x') \right\} =
\left\{ \widetilde{\Omega}_{12}(x),\widetilde{\Omega}_{12}(x')
\right\} = \left\{ \widetilde{\Omega}(x) \mathop{,}^{\otimes}
\widetilde{\Omega}(x') \right\}_{12,12} = 0.
\end{equation*}

\section{Dirac Brackets of the action-angle variables} \label{section:
Dirac brackets}

In order to isolate the action-angle variables, i.e. the physical
degrees of freedom of the string, we imposed the Virasoro
constraints and static gauge fixing condition on the reconstructed
current. However, these constraints together form a set of second
class constraints. Therefore the algebra of the action-angle
variables should be expressed in terms of Dirac brackets instead
of Poisson brackets. In this appendix we show that the Dirac
brackets of the action-angle variables are (weakly\footnote{In the
context of constrained Hamiltonian systems, two functions on
phase-space are said to be `weakly' equal if they are equal on the
constraint surface. Note that this concept of weakness bears no
relation to the notion of a `weak' Poisson bracket introduced in
section \ref{section: Maillet}.}) equal to their Poisson brackets.

We start with the worldsheet action for a string on $\mathbb{R}
\times S^3$ in conformal gauge. It is possible to work in
conformal gauge right from the outset since the worldsheet metric
and its conjugate momentum form a pair of second class constraints
that commutes with the Virasoro constraints. The $S^3$ and
$\mathbb{R}$ parts of the action decouple with the equations of
motion for the $\mathbb{R}$ part being
\begin{equation*}
d \ast d X_0 = 0.
\end{equation*}
In conformal gauge this reads $\partial_+ \partial_- X_0 = 0$ which
admits the general solution
\begin{equation*}
X_0(\sigma,\tau) = X_0^+(\sigma^+) + X_0^-(\sigma^-).
\end{equation*}
The equations of motion for the $S^3$ part $d \ast j = 0, dj - j
\wedge j = 0$ or equivalently
\begin{equation} \label{eom for j}
\partial_- j_+ = - \partial_+ j_- = - \frac{1}{2} [j_+, j_-],
\end{equation}
can be rewritten as a zero curvature condition for a Lax
connection
\begin{equation} \label{zero curvature}
dJ(x) - J(x) \wedge J(x) = 0, \qquad J(x) = \frac{j - x \ast j}{1
- x^2} \in \mathfrak{sl}(2,\mathbb{C}).
\end{equation}
Using this flat connection we can define an algebraic curve
$\Sigma$ in $\mathbb{C}^2$ as a desingularisation of the spectral
curve
\begin{equation*}
\Gamma : \quad \Gamma(x,y) = \text{det}\left( y {\bf 1} -
\Omega(x,\sigma,\tau) \right) = 0, \qquad \Omega(x,\sigma,\tau) \equiv
P \overleftarrow{\exp} \int_{[c(\sigma,\tau)]} J(x) \in SL(2,\mathbb{C}).
\end{equation*}
As in section \ref{section: BA vector} the general solution is
reconstructed by identifying the analytic properties of the
Baker-Akhiezer vector $\bm{\psi}(P), P \in \Sigma$ which solves
the auxiliary linear system for which \eqref{zero curvature} is
the consistency condition
\begin{equation} \label{auxiliary}
\left( d - J(x) \right) \bm{\psi} = 0.
\end{equation}
In order to compute the Dirac brackets one must relax the Virasoro
constraints and static gauge fixing condition in the
reconstruction. One then finds that $\bm{\psi}$ is uniquely determined
by
\begin{gather*}
(\psi_1) \geq \hat{\gamma}^{-1} \infty^-, \quad \psi_1(\infty^+) = 1,
\quad \text{and} \quad (\psi_2) \geq \hat{\gamma}^{-1} \infty^+, \quad
\psi_2(\infty^-) = 1,\\
\text{with} \quad \left\{ \begin{array}{l}
\psi_i(x^{\pm},\sigma,\tau) \exp \left( \mp \frac{f_+(\sigma^+)}{1
- x} \right) = O(1), \quad \text{as} \; x \rightarrow 1,\\
\psi_i(x^{\pm},\sigma,\tau) \exp \left( \mp \frac{f_-(\sigma^-)}{1
+ x} \right) = O(1), \quad \text{as} \; x \rightarrow
-1,\end{array} \right.
\end{gather*}
where $f_{\pm}$ are two arbitrary functions related to the
conformal invariance of the equations of motion \eqref{eom for j}.
Explicit reconstruction requires the introduction of an Abelian
differential $d\mathcal{Q}$ of the second kind on $\Sigma$ defined
by its pole structure at $x = \pm 1$, namely
\begin{equation*}
d\mathcal{Q}(x^{\pm}) \underset{x \rightarrow + 1}\sim \pm
f_+(\sigma^+) \frac{dx}{(1 - x)^2}, \qquad d\mathcal{Q}(x^{\pm})
\underset{x \rightarrow - 1}\sim \pm f_-(\sigma^-) \frac{dx}{(1 +
x)^2}.
\end{equation*}
We can write $d\mathcal{Q} = f_+(\sigma^+) dp_+ + f_-(\sigma^-) dp_-$
where $dp_{\pm}$ are Abelian differentials of the second kind defined
by their respective poles at $x = \pm 1$,
\begin{equation*}
dp_+(x^{\pm}) \underset{x \rightarrow +1}\sim \pm \frac{dx}{(1 -
x)^2}, \quad dp_-(x^{\pm}) \underset{x \rightarrow -1}\sim \pm
\frac{dx}{(1 + x)^2}, \quad dp_{\pm} \underset{x \rightarrow \mp
1}\sim O\left((1 \pm x)^0\right).
\end{equation*}

Just as in the flat space case, here the general solution to the
equations of motion is a function of $\sigma^{\pm}$, through the
differential $d\mathcal{Q} = f_+(\sigma^+) dp_+ + f_-(\sigma^-)
dp_-$, which is what we expect since the equations of motion for
the current $j$ are conformally invariant, being derived from a
conformally invariant action. So we have the following general
solution for the sting moving on $\mathbb{R} \times S^3$ in
conformal gauge
\begin{equation} \label{general sol}
X^{\text{sol}}_0(\sigma,\tau) = X_0^+(\sigma^+) + X_0^-(\sigma^-) \in
\mathbb{R},\qquad j^{\text{sol}}(\sigma,\tau) = j\left(f_+(\sigma^+),
f_-(\sigma^-)\right) \in SU(2),
\end{equation}
where $X_0^{\pm}, f_{\pm}$ are arbitrary functions. We note here that
the effect of the Virasoro constraint is to relate these arbitrary
functions, precisely we have
\begin{equation*}
\frac{1}{2} \text{tr} j_{\pm}^2 + (\partial_{\pm} X_0)^2 = 0
\quad \Leftrightarrow \quad f_{\pm}(\sigma^{\pm}) =
X_0^{\pm}(\sigma^{\pm}).
\end{equation*}
The effect of the static gauge fixing condition on the other hand
is to fix completely the arbitrariness of the functions $X_0^{\pm}$,
namely
\begin{equation*}
X_0 = \kappa \tau \quad \Leftrightarrow \quad X_0^{\pm}(\sigma^{\pm})
= \frac{\kappa}{2} \sigma^{\pm}.
\end{equation*}
We make use of the general solutions \eqref{general sol} to
parameterise the phase space variables as follows
\begin{equation*}
\left( X_0(\sigma), \Pi_0(\sigma), j_{\pm}(\sigma) \right) = \left(
X^{\text{sol}}_0(\sigma,0), \partial_{\tau}
X^{\text{sol}}_0(\sigma,0), j^{\text{sol}}_{\pm}(\sigma,0) \right).
\end{equation*}

We now need to impose the Virasoro constraints on phase-space
\begin{subequations} \label{constraints}
\begin{equation} \label{Virasoro constraints}
T_{\pm \pm} \equiv \frac{1}{2} \text{tr} j^2_{\pm} + \left( \frac{2
\pi}{\sqrt{\lambda}} \Pi_0 \mp \partial_{\sigma} X_0 \right)^2 \approx
0,
\end{equation}
as well as get rid of the residual gauge (i.e. conformal)
invariance by imposing a further gauge fixing condition, which we
choose to be the static gauge\footnote{One can use the residual gauge
freedom $\sigma^{\pm} \rightarrow \tilde{\sigma}^{\pm} =
h_{\pm}(\sigma^{\pm})$ to set $\tilde{\tau} \propto X_0$ since
$\tilde{\tau} = \frac{1}{2}\left( h_+(\sigma^+) + h_-(\sigma^-)
\right)$ solves the equations of motion for $X_0$. The coefficient of
proportionality is forced on us by conformal invariance
of the quantity $p_0 = \int_0^{2 \pi} d\sigma \Pi_0(\sigma,\tau) = -
\frac{\sqrt{\lambda}}{2 \pi} \int_0^{2 \pi} d\sigma
\dot{X}_0(\sigma,\tau)$, so that $X_0 = - \frac{p_0}{\sqrt{\lambda}}
\tilde{\tau}$.}
\begin{equation} \label{Static gauge}
X_0 \approx - \frac{p_0}{\sqrt{\lambda}} \tau, \quad \Pi_0 \approx
\frac{p_0}{2 \pi},
\end{equation}
\end{subequations}
where $p_0$ is the zero mode of $\Pi_0$. One can show that
\begin{equation} \label{Vir algebra}
\left\{ \frac{1}{2} \text{tr} j_{\pm}^2 (\sigma), \frac{1}{2}
\text{tr} j_{\pm}^2 (\sigma') \right\} = \pm \frac{8
\pi}{\sqrt{\lambda}} \left[ \frac{1}{2} \text{tr} j_{\pm}^2 (\sigma) +
\frac{1}{2} \text{tr} j_{\pm}^2 (\sigma') \right] \delta'(\sigma -
\sigma').
\end{equation}
and likewise $( \frac{2 \pi}{\sqrt{\lambda}} \Pi_0 \mp
\partial_{\sigma} X_0 )^2$ satisfies the same equation, so that the
Virasoro constraints $T_{\pm \pm}$ by themselves are first
class. However, the static gauge constraints fail to commute with
these and among themselves (since $\{ \Pi_0(\sigma), X_0(\sigma') \} =
\delta(\sigma - \sigma') \not \approx 0$), and so the constraints in
\eqref{constraints} are second class. In terms of modes, the
constraints \eqref{constraints} read
\begin{equation*}
\alpha_n \approx \tilde{\alpha}_n \approx 0, \quad x_0 + \frac{p_0}{\sqrt{\lambda}} \tau
\approx 0, \qquad L_n \approx \tilde{L}_n \approx 0, \quad L_0 \approx
\tilde{L}_0 \approx -\frac{p^2_0}{4 \sqrt{\lambda}}, \quad n \neq 0,
\end{equation*}
where $L_n, \tilde{L}_n$ are the fourier modes of $\frac{1}{2}
\text{tr} j^2_{\pm}$ respectively,
\begin{equation} \label{Virasoro modes}
L_n = \frac{\sqrt{\lambda}}{8 \pi} \int_0^{2 \pi} d\sigma e^{i n \sigma} \frac{1}{2}
\text{tr}j_+^2(\sigma), \qquad \tilde{L}_n = \frac{\sqrt{\lambda}}{8
\pi} \int_0^{2 \pi} d\sigma e^{- i n \sigma} \frac{1}{2}
\text{tr}j_-^2(\sigma),
\end{equation}
satisfying the following algebra,
\begin{equation*}
\begin{split}
\{ L_m, L_n \} &= i (n - m) L_{m + n},\\
\{ L_m, \tilde{L}_n \} &= 0,\\
\{ \tilde{L}_m, \tilde{L}_n \} &= i (n - m) \tilde{L}_{m +
n},
\end{split}
\end{equation*}
which follows from \eqref{Vir algebra}, and $\alpha_n,
\tilde{\alpha}_n$ are the modes of $X_0$ and $\Pi_0$ defined by
\begin{equation} \label{extract data: flat}
\begin{split}
&\alpha_n = \frac{\lambda^{\frac{1}{4}}}{\sqrt{2} \pi} \int_0^{2 \pi} d\sigma e^{-i n \sigma}
\frac{1}{2} \left( - \frac{2 \pi}{\sqrt{\lambda}} \Pi_0(\sigma) -
\partial_{\sigma} X_0(\sigma) \right), \quad n \neq 0 \\
&\tilde{\alpha}_n = \frac{\lambda^{\frac{1}{4}}}{\sqrt{2} \pi}
\int_0^{2 \pi} d\sigma e^{i n \sigma} \frac{1}{2} \left( - \frac{2
\pi}{\sqrt{\lambda}} \Pi_0(\sigma) + \partial_{\sigma} X_0(\sigma)
\right), \quad n \neq 0 \\ &x_0 = \int_0^{2 \pi} d\sigma X_0(\sigma), \qquad
p_0 = \int_0^{2 \pi} d\sigma \Pi_0(\sigma),
\end{split}
\end{equation}
satisfying the following algebra,
\begin{equation*}
\begin{split}
\{ \alpha_m, \alpha_n \} &= i m \delta_{m + n}, \quad \{ \alpha_m,
\tilde{\alpha}_n \} = 0,\\ \{ \tilde{\alpha}_m, \tilde{\alpha}_n \} &=
i m \delta_{m + n}, \quad \{ p_0, x_0 \} = 1.
\end{split}
\end{equation*}

For the closed string, static gauge does not completely fix the
residual gauge invariance as there still remains the possibility
of rigid translations $\sigma \rightarrow \sigma + b$, which is
generated by $L_0 - \tilde{L}_0$. This rigid transformation can be
dealt with by symplectic reduction as explained in \cite{Paper1},
which consists in imposing the constraint that the total worldsheet
momentum vanishes $\mathcal{P} \propto  L_0 - \tilde{L}_0 \propto
-\sum_{I=1}^K n_I \mathcal{S}_I = 0$ as well as identifying points
related by translations in $\sigma$. So setting aside this rigid
transformation, the set of relevant constraints thus reads
\begin{equation} \label{constraints 2}
\alpha_n \approx \tilde{\alpha}_n \approx 0, \quad x_0 +
\frac{p_0}{\sqrt{\lambda}} \tau
\approx 0, \qquad L_n \approx \tilde{L}_n \approx 0, \quad L_0 +
\tilde{L}_0 + \frac{p^2_0}{2 \sqrt{\lambda}} \approx 0, \quad n \neq 0.
\end{equation}
In order to fix these constraints one must replace the Poisson
bracket by the Dirac bracket for this set of second class
constraints. The question now is whether the Dirac bracket for the
action angle variables are the same as their Poisson brackets.

Now for functions $f,g$ of the principal chiral fields $j$ (which are
independent of $X_0, \Pi_0$ and therefore commute with the constraints
$\alpha_n, \tilde{\alpha}_n, x_0 + p_0 \tau/\sqrt{\lambda}$) the Dirac bracket takes
the schematic form
\begin{equation*}
\begin{split}
\{ f, g \}_D = \{ f, g \} &+ \{ f, L_n \} A_{nm} \{ L_m, g \} + \{
f, L_0 + \tilde{L}_0 \} B_m \{ L_m, g \} + \{ f, L_m \} C_m \{ L_0
+ \tilde{L}_0, g \}\\ &+ \{ f, \tilde{L}_n \} \tilde{A}_{nm} \{
\tilde{L}_m, g \} + \{ f, L_0 + \tilde{L}_0 \} \tilde{B}_m \{
\tilde{L}_m, g \} + \{ f, \tilde{L}_m \} \tilde{C}_m \{ L_0 +
\tilde{L}_0, g \}.
\end{split}
\end{equation*}
Note that there is no term of the form $``\{ f, L_0 + \tilde{L}_0
\} D \{ L_0 + \tilde{L}_0, g \}"$ because the corresponding
component $D$ in the inverse matrix $\mathcal{C}_{ab}^{-1} = \{
\phi_a, \phi_b \}^{-1}$ of the Poisson bracket of constraints
vanishes. This property boils down to the fact that the constraint
$x_0 + p_0 \tau/\sqrt{\lambda}$ commutes with every constraint in
\eqref{constraints 2} including itself but only fails to commute with
the constraint $L_0 + \tilde{L}_0 + p_0^2 / (2 \sqrt{\lambda})$,
\begin{equation*}
\begin{array}{rl}
\mathcal{C}_{ab} = \{ \phi_a, \phi_b \} = \left(
\begin{array}{c|ccccc}
0 & \ast & \ast & 0 & 0 & \ast\\
\hline
\ast & \ast & 0 & 0 & 0 & 0\\
\ast & 0 & \ast & 0 & 0 & 0\\
0 & 0 & 0 & \ast & 0 & 0\\
0 & 0 & 0 & 0 & \ast & 0\\
\ast & 0 & 0 & 0 & 0 & 0
\end{array}
\right) & \begin{array}{l} L_0 + \tilde{L}_0 + p_0^2 / (2
\sqrt{\lambda})\\ L_n\\ \tilde{L}_n\\ \alpha_n\\ \tilde{\alpha}_n\\
x_0 + p_0 \tau / \sqrt{\lambda} \end{array}.
\end{array}
\end{equation*}
It follows that for functions $f,g$ of the principal chiral model
that are invariant under residual gauge transformations generated
by $L_n, \tilde{L}_n, n \neq 0$ we have the desired equality of
Dirac and Poisson brackets
\begin{equation*}
\{ f, g \}_D = \{ f, g \}.
\end{equation*}
It therefore remains to check that the action angle variables can
be defined in a conformally invariant way from the general
solution $j^{\text{sol}}(\sigma, \tau)$ obtained in \eqref{general
sol}.

Going back to expression \eqref{general sol} we see that the
periodicity requirement of the solution under $\sigma \rightarrow
\sigma + 2 \pi$ leads to the conditions
\begin{equation*}
\left[ f_{\pm}(\sigma + 2 \pi) - f_{\pm}(\sigma) \right] \int_{\bm{b}}
dp_{\pm} \in 2 \pi \mathbb{Z}^g + 2 \pi \Pi \mathbb{Z}^g.
\end{equation*}
Since the Jacobian lattice $2 \pi \mathbb{Z}^g + 2 \pi \Pi
\mathbb{Z}^g$ is discrete, this in turn requires the expression in
square brackets to be constant, i.e. independent of $\sigma$, say
$f_{\pm}(\sigma + 2 \pi) - f_{\pm}(\sigma) = 2 \pi k^{\pm}$, $k^{\pm}
\in \mathbb{C}$. Then $f_{\pm}(\sigma) - k^{\pm} \sigma$ is periodic
under $\sigma \rightarrow \sigma + 2 \pi$, which means that we can
decompose the functions $f_{\pm}$ as follows
\begin{equation} \label{function f}
f_{\pm}(\sigma) = \xi^{\pm}_0 + k^{\pm} \sigma + \sum_{n \neq 0} \xi^{\pm}_n e^{i n
\sigma}.
\end{equation}
Recall that imposing the Virasoro constraint on the solution
$j^{\text{sol}}(\sigma,\tau)$ has the effect of rendering the
functions $f_{\pm}$ linear, and so this corresponds to setting all the
modes $\xi^{\pm}_n$ in \eqref{function f} to zero, i.e.
\begin{equation} \label{constraint surface}
\frac{1}{2} \text{tr} \left(j^{\text{sol}}_{\pm}\right)^2 = -
\frac{p^2_0}{\lambda} \qquad \Leftrightarrow \qquad \xi^{\pm}_n = 0,
\; \forall n, \quad k^{\pm} = \frac{i p_0}{2 \sqrt{\lambda}}.
\end{equation}

As we now show, the effect of the Virasoro constraints on the
functions $f_{\pm}$ can be deduced from the following brackets
\begin{equation*}
\begin{split}
\left\{ \frac{\sqrt{\lambda}}{4 \pi} \frac{1}{2} \text{tr}
j_{\pm}^2(\sigma), j_{\pm}^b(\sigma') \right\} &= \frac{1}{2} \left[
j_{\pm}(\sigma), j_{\mp}(\sigma) \right]^b \delta(\sigma - \sigma')
\pm 2 j_{\pm}^b(\sigma) \delta'(\sigma - \sigma'), \\
\left\{ \frac{\sqrt{\lambda}}{4 \pi} \frac{1}{2} \text{tr}
j_{\pm}^2(\sigma), j_{\mp}^b(\sigma') \right\} &= - \frac{1}{2} \left[
j_{\pm}(\sigma), j_{\mp}(\sigma) \right]^b \delta(\sigma - \sigma'),
\end{split}
\end{equation*}
which are a consequence of the non-ultra local brackets of the
principal chiral model. Let $j^{\text{sol}}(\sigma,\tau)$ be a
physical path, i.e. satisfying the equations of motion \eqref{eom for
j}, then one can deduce immediately from the above brackets that
\begin{equation*}
\begin{split}
\left\{ \frac{\sqrt{\lambda}}{4 \pi} \int \epsilon^{\pm}(\sigma' \pm
\tau) \frac{1}{2} \text{tr} j_{\pm}^2(\sigma') d\sigma',
j_{\pm}^b(\sigma) \right\} \left( j^{\text{sol}}(\sigma,\tau) \right)
&= - \partial_{\pm} \left( \epsilon^{\pm}(\sigma \pm \tau)
j_{\pm}^{\text{sol}}(\sigma,\tau) \right)^b,\\
\left\{ \frac{\sqrt{\lambda}}{4 \pi} \int \epsilon^{\pm}(\sigma' \pm \tau)
\frac{1}{2} \text{tr} j_{\pm}^2(\sigma') d\sigma', j_{\mp}^b(\sigma) \right\}
\left( j^{\text{sol}}(\sigma,\tau) \right) &= - \epsilon^{\pm}(\sigma \pm
\tau) \left( \partial_{\pm} j_{\mp}^{\text{sol}}(\sigma,\tau) \right)^b.
\end{split}
\end{equation*}
So $j^b_{\pm}$ transforms as a scalar under $\frac{1}{2} \text{tr}
j_{\mp}^2$ but as a scalar density of weight 1 under $\frac{1}{2} \text{tr}
j_{\pm}^2$ (this is in agreement with the fact that the Langrangian
$L \propto \text{tr} (j_+ j_-)$ should be a density of weight
1 under coordinate transformations). Because
$j^{\text{sol}}(\sigma,\tau) = j(f_+(\sigma^+), f_-(\sigma^-))$ one
can now derive the action of $L_n, \tilde{L}_n$ on the functions
$f_{\pm}$, namely
\begin{equation*}
\begin{split}
\{ L_n, f_+(\sigma^+) \} &= - e^{i n \sigma^+} \partial_{\sigma^+}
f_+(\sigma^+), \qquad \quad \{ L_n, f_-(\sigma^-) \} = - e^{i n \sigma^+} \partial_{\sigma^+}
f_-(\sigma^-) = 0,\\
\{ \tilde{L}_n, f_+(\sigma^+) \} &= - e^{-i n \sigma^-} \partial_{\sigma^-}
f_+(\sigma^+) = 0, \qquad \{ \tilde{L}_n, f_-(\sigma^-) \} = - e^{-i n \sigma^-} \partial_{\sigma^-}
f_-(\sigma^-).
\end{split}
\end{equation*}

Using the definition \eqref{function f} of the functions $f_{\pm}$ we
are now able to write the action of the Virasoro constraints $L_n,
\tilde{L}_n, n \neq 0$ on the parameters $\xi^+_m$ and
$\xi^-_m$. Explicitly we find
\begin{equation*}
\begin{split}
\{ L_n, \xi^-_m \} = 0, \qquad \{ L_n, \xi^+_m \} = \delta_{m n} k^+ -
(m - n) \xi^+_{m - n},\\
\{ \tilde{L}_n, \xi^+_m \} = 0, \qquad \{ \tilde{L}_n, \xi^-_m \} =
\delta_{m n} k^- - (m - n) \xi^-_{m - n}.
\end{split}
\end{equation*}
In particular, on the constraint surface \eqref{constraint surface} we get
\begin{equation} \label{conf tr of xi}
\{ L_n, \xi^+_m \} \approx \delta_{m n} k^+, \qquad \{ \tilde{L}_n, \xi^-_m
\} \approx \delta_{m n} k^-, \quad n \neq 0.
\end{equation}

Recalling how the $\sigma, \tau$ evolution of the solution is
expressed in terms of the Abelian differential of the second kind $d
\mathcal{Q} = f_+(\sigma^+) dp_+ + f_-(\sigma^-) dp_-$ as
\begin{equation*}
\begin{split}
\bm{\theta}(\sigma,\tau) = \bm{\theta}_0 - \int_{\bm{b}} d\mathcal{Q}
&= \bm{\theta}_0 - f_+(\sigma^+) \int_{\bm{b}} dp_+ - f_-(\sigma^-)
\int_{\bm{b}} dp_-,\\
\bar{\theta}(\sigma,\tau) = \bar{\theta}_0 -
\int_{\infty^-}^{\infty^+} d\mathcal{Q} &= \bar{\theta}_0 -
f_+(\sigma^+) \int_{\infty^-}^{\infty^+} dp_+ - f_-(\sigma^-)
\int_{\infty^-}^{\infty^+} dp_-,
\end{split}
\end{equation*}
the angle variables were defined in \cite{Paper1} and in section
\ref{section: action-angle} simply as the parameters
$\bm{\theta}(0,0), \bar{\theta}(0,0)$. A more suitable definition
here, valid off the constraint surface, would be instead to take the
angle variable $\varphi_I,I = 1, \ldots, K = g+1$ as the zero mode of
$\theta_i(\sigma,0), \bar{\theta}(\sigma,0)$, namely on $J(\Sigma)
\times \mathbb{C}/2 \pi \mathbb{Z}$ we define 
\begin{equation} \label{angle variables alternative}
\begin{split}
\bm{\varphi} &= \bm{\theta}_0 - \xi_0^+ \int_{\bm{b}} dp_+ - \xi_0^-
\int_{\bm{b}} dp_-,\\
\varphi_K &= \bar{\theta}_0 - \xi_0^+ \int_{\infty^-}^{\infty^+} dp_+ -
\xi_0^- \int_{\infty^-}^{\infty^+} dp_-.
\end{split}
\end{equation}
The difference between these two definitions is the following vector
in $J(\Sigma) \times \mathbb{C}/2 \pi \mathbb{Z}$
\begin{equation*}
\begin{split}
\bm{\theta}(0,0) - \bm{\varphi} &= - \left( \sum_{n \neq 0} \xi_n^+ \right)
\int_{\bm{b}} dp_+ - \left( \sum_{n \neq 0} \xi_n^- \right)
\int_{\bm{b}} dp_- \approx 0,\\
\bar{\theta}(0,0) - \varphi_K &= - \left( \sum_{n \neq 0} \xi_n^+ \right)
\int_{\infty^-}^{\infty^+} dp_+ - \left( \sum_{n \neq 0} \xi_n^-
\right) \int_{\infty^-}^{\infty^+} dp_- \approx 0,
\end{split}
\end{equation*}
which vanishes on the constraint surface. In particular, on the constraint
surface \eqref{constraint surface} we have by \eqref{conf tr of xi}
and \eqref{angle variables alternative}
\begin{equation*}
\{ L_n, \varphi_I \} \approx \{ \tilde{L}_n, \varphi_I \} \approx 0, \quad
I = 1, \ldots, K.
\end{equation*}
Since the action variables $S_I, I= 1,\ldots, K$ are invariant
under $\sigma,\tau$ evolution, they obviously Poisson commute with the
generators of conformal transformation $L_n, \tilde{L}_n$ and so we
also have
\begin{equation*}
\{ L_n, S_I \} = \{ \tilde{L}_n, S_I \} = 0, \quad I = 1, \ldots, K.
\end{equation*}

So finally we have established equality of the Dirac and Poisson
bracket of the action angle variables on the constraint surface,
\begin{equation*}
\{ f, g \}_D \approx \{ f, g \}, \qquad \text{for} \; f,g \in \{
\varphi_I, S_I \}_{I = 1}^K.
\end{equation*}


\begin{thebibliography}{5}
\bibitem{Paper1}
  N.~Dorey and B.~Vicedo,
  ``On the dynamics of finite-gap solutions in classical string theory,''
  arXiv:hep-th/0601194.

\bibitem{Faddeev:1987ph}
  L.~D.~Faddeev and L.~A.~Takhtajan,
  \textit{Hamiltonian Methods In The Theory Of Solitons},
  Springer-Verlag Berlin (1987)

\bibitem{Krichever:2001zg}
  I.~Krichever,
  ``Vector bundles and Lax equations on algebraic curves,''
  Commun.\ Math.\ Phys.\  {\bf 229} (2002) 229,
  arXiv:hep-th/0108110.

\bibitem{Metsaev:1998it}
  R.~R.~Metsaev and A.~A.~Tseytlin,
  ``Type IIB superstring action in AdS(5) x S(5) background,''
  Nucl.\ Phys.\ B {\bf 533} (1998) 109,
  arXiv:hep-th/9805028.

\bibitem{Bena:2003wd}
  I.~Bena, J.~Polchinski and R.~Roiban,
  ``Hidden symmetries of the AdS(5) x S**5 superstring,''
  Phys.\ Rev.\ D {\bf 69} (2004) 046002,
  arXiv:hep-th/0305116.

\bibitem{Kazakov:2004qf}
  V.~A.~Kazakov, A.~Marshakov, J.~A.~Minahan and K.~Zarembo,
  ``Classical / quantum integrability in AdS/CFT,''
  JHEP {\bf 0405} (2004) 024,
  arXiv:hep-th/0402207.

\bibitem{Beisert:2004ag}
  N.~Beisert, V.~A.~Kazakov and K.~Sakai,
  ``Algebraic curve for the SO(6) sector of AdS/CFT,''
  arXiv:hep-th/0410253.

\bibitem{Beisert:2005bm}
  N.~Beisert, V.~A.~Kazakov, K.~Sakai and K.~Zarembo,
  ``The algebraic curve of classical superstrings on AdS(5) x S**5,''
  arXiv:hep-th/0502226.

 \bibitem{Schafer-Nameki:2004ik}
   S.~Schafer-Nameki,
   ``The algebraic curve of 1-loop planar N = 4 SYM,''
   Nucl.\ Phys.\ B {\bf 714} (2005) 3,
   arXiv:hep-th/0412254.

 \bibitem{Alday:2005gi}
   L.~F.~Alday, G.~Arutyunov and A.~A.~Tseytlin,
   ``On integrability of classical superstrings in AdS(5) x S**5,''
   JHEP {\bf 0507} (2005) 002,
   arXiv:hep-th/0502240.

\bibitem{Babelon}
  O.~Babelon, D.~Bernard, M.~Talon,
  \textit{Introduction to Classical Integrable Systems},
  Cambridge University Press (2003)

\bibitem{GSWI} M.~B.~Green, J.~H.~Schwarz and E.~Witten,
  ``Superstring Theory. Vol. 1: Introduction,''
\bibitem{BrH}
  L.~Brink and M.~Henneaux,
  ``Principles Of String Theory,''

\bibitem{Krichever1}
  I.~M.~Krichever,
  ``Integration of non-linear equations by methods of algebraic geometry,''
  Funct. Anal. Appl. {\bf 11} (1) (1977), 12-26

\bibitem{Krichever2}
  I.~M.~Krichever,
  ``Methods of algebraic geometry in the theory of non-linear equations,''
  Russian Math. Surveys {\bf 32} (6) (1977), 185-213

\bibitem{Krichever+Phong}
  I.~M.~Krichever and D.~H.~Phong,
  ``On the integrable geometry of soliton equations and N = 2  supersymmetric
  gauge theories,''
  J.\ Diff.\ Geom.\  {\bf 45} (1997) 349,
  arXiv:hep-th/9604199.

\bibitem{Krichever3} I.~M.~Krichever,  ``Two-dimensional
  algebraic-geometrical operators with self-consistent potentials'',
  Func. An \& Apps. {\bf 28} (1994) No 1, 26.

\bibitem{Belokolos}
  E.~D.~Belokolos, A.~I.~Bobenko, V.~Z.~Enol'skii, A.~R.~Its, V.~B.~Matveev,
  \textit{Algebro-Geometric Approach to Nonlinear Integrable Equations},
  Springer-Verlag Telos (1994)

\bibitem{Sklyanin:1995bm}
  E.~K.~Sklyanin,
  ``Separation of variables - new trends,''
  Prog.\ Theor.\ Phys.\ Suppl.\  {\bf 118} (1995) 35.

\bibitem{Maillet2}
  J.~M.~Maillet,
  ``Kac-Moody algebra and extended Yang-Baxter relations in the O(N)
  non-linear $\sigma$-model'',
  Phys. Lett. {\bf 162}B, 137-142 (1985)

\bibitem{Maillet3}
  J.~M.~Maillet,
  ``Hamiltonian Structures For Integrable Classical Theories From
  Graded Kac-Moody Algebras,''
  Phys. Lett. B {\bf 167} (1986) 401.

\bibitem{Maillet}
  J.~M.~Maillet,
  ``New integrable canonical structures in two-dimensional models'',
  Nucl. Phys. B {\bf 269}, 54-76 (1986)

\bibitem{BFLS}
  M.~Bordemann, M.~Forger, J.~Laartz and U.~Sch$\ddot{\text{a}}$per,
  ``The Lie-Poisson Structure of Integrable Classical Non-Linear Sigma Model'',
  Commun. Math. Phys. {\bf 152}, 167-190 (1993)

\bibitem{Fedorov}
  Y.~Fedorov,
  ``Classical Integrable Systems and Billiards Related to Generalized
  Jacobians'',
  Acta Appl. Math. {\bf 55}, 3, 151-201 (1999)

\end{thebibliography}
\end{document}